\documentclass[trackchanges,twocolumn,twocolappendix]{aastex701}

\newcommand{\Prot}{$17.33^{+0.29}_{-0.28}$}

\begin{document}

\title{The First Adolescent Super-Earth-Sized Multi-Planet System Around an M Dwarf: TOI-2079}

\author[orcid=0000-0001-9204-8498,sname='Peláez-Torres']{Alberto Peláez-Torres}
\affiliation{Instituto de Astrofísica de Andalucía (IAA-CSIC), Gta. de la Astronomía s/n, 18008 Granada, Spain}
\email{apelaez@iaa.es}

\author[orcid=0000-0003-1572-7707,sname='Pozuelos']{Francisco J. Pozuelos}
\affiliation{Instituto de Astrofísica de Andalucía (IAA-CSIC), Gta. de la Astronomía s/n, 18008 Granada, Spain}
\email{pozuelos@iaa.es}

\author[orcid=0000-0002-2341-3233,sname='Esparza-Borges']{Emma Esparza-Borges}
\affiliation{European Space Agency (ESA), European Space Astronomy Centre (ESAC), Camino Bajo del Castillo s/n, 28692 Villanueva de
la Ca{\~n}ada, Madrid, Spain}
\email{emma.esparza.borges@gmail.com}

\author[orcid=0000-0002-4262-5661,sname='Morello']{Giuseppe Morello}
\affiliation{Instituto de Astrofísica de Andalucía (IAA-CSIC), Gta. de la Astronomía s/n, 18008 Granada, Spain}
\affiliation{INAF- Palermo Astronomical Observatory, Piazza del Parlamento, 1, 90134 Palermo, Italy}
\email{gmorello@iaa.es}

\author[orcid=0000-0003-2007-6144,sname='Dévora-Pajares']{Martín Dévora-Pajares}
\affiliation{Dpto. Física Teórica y del Cosmos, Universidad de Granada, 18071 Granada, Spain}
\email{martin.devora.pajares@gmail.com} 

\author[orcid=0000-0002-1397-8169,sname='Gkouvelis']{Leonardos Gkouvelis}
\affiliation{Instituto de Astrofísica de Andalucía (IAA-CSIC), Gta. de la Astronomía s/n, 18008 Granada, Spain}
\email{Leo.Gkouvelis@physik.uni-muenchen.de}

\author[orcid=0000-0003-2527-1475,sname='Vissapragada']{Shreyas Vissapragada}
\affiliation{The Observatories of the Carnegie Institution for Science, 813 Santa Barbara Street, Pasadena, CA 91101, USA}
\email{svissapragada@carnegiescience.edu} 

\author[orcid=0000-0002-0149-1302,sname='Rojas Ayala']{B\'arbara Rojas-Ayala}
\affiliation{Instituto de Alta Investigaci\'on, Universidad de Tarapac\'a, Casilla 7D, Arica, Chile}
\email{brojasayala@academicos.uta.cl} 

\author[orcid=0000-0002-6523-9536,sname='Burgasser']{Adam J.\ Burgasser}
\affiliation{Department of Astronomy \& Astrophysics, University of California, San Diego, La Jolla, CA, USA}
\email{aburgasser@ucsd.edu} 

\author[orcid=0000-0002-1397-8169,sname='Rackham']{Benjamin V. Rackham}
\affiliation{Department of Earth, Atmospheric and Planetary Sciences, Massachusetts Institute of Technology, Cambridge, MA 02139, USA}
\affiliation{Kavli Institute for Astrophysics and Space Research, Massachusetts Institute of Technology, Cambridge, MA 02139, USA}
\email{brackham@mit.edu} 

\author[orcid=0000-0003-2171-5083,sname='Tamburo']{Patrick Tamburo}
\affiliation{Center for Astrophysics \textbar\ Harvard \& Smithsonian, 60 Garden Street, Cambridge, MA 02138, USA}
\email{patrick.tamburo@cfa.harvard.edu}

\author[orcid=0000-0003-2066-8959,sname='Orell-Miquel']{Jaume Orell-Miquel}
\affiliation{Department of Astronomy, University of Texas at Austin, Austin, TX, USA}
\email{jaume.miquel@austin.utexas.edu} 

\author[orcid=0000-0003-0987-1593,sname='Pallé']{Enric Pallé}
\affiliation{Instituto de Astrof\'isica de Canarias (IAC), E-38200 La Laguna, Tenerife, Spain}
\affiliation{Departamento de Astrof\'isica, Universidad de La Laguna (ULL), E-38206 La Laguna, Tenerife, Spain}
\email{epalle@iac.es}

\author[orcid=,sname='Narita']{Norio Narita}
\affiliation{Komaba Institute for Science, The University of Tokyo, 3-8-1 Komaba, Meguro, Tokyo 153-8902, Japan}
\affiliation{Astrobiology Center, 2-21-1 Osawa, Mitaka, Tokyo 181-8588, Japan}
\affiliation{Instituto de Astrof\'isica de Canarias (IAC), E-38200 La Laguna, Tenerife, Spain}
\email{narita@g.ecc.u-tokyo.ac.jp}

\author[orcid=0000-0001-6588-9574,sname='Collins']{Karen A.\ Collins}
\affiliation{Center for Astrophysics \textbar \ Harvard \& Smithsonian, 60 Garden Street, Cambridge, MA 02138, USA}
\email{karen.collins@cfa.harvard.edu}

\author[orcid=0009-0002-5067-5463,sname='Abreu']{Nestor Abreu Garcia}
\affiliation{Instituto de Astrof\'isica de Canarias (IAC), E-38200 La Laguna, Tenerife, Spain}
\affiliation{Departamento de Astrof\'isica, Universidad de La Laguna (ULL), E-38206 La Laguna, Tenerife, Spain}
\email{alu0100996609@ull.edu.es} 

\author[orcid=,sname='Akar']{Furkan Akar}
\affiliation{Ankara University, Graduate School of Natural \& Applied Sciences, Astronomy \& Space Sciences Department, Ziraat Mahallesi {\.I}rfan Ba\c{s}tu\u{g} Caddesi, D{\i}\c{s}kap{\i}, TR-06110 Alt{\i}nda\u{g}, Ankara, T\"urkiye}
\email{furkanakar_@outlook.com}

\author[orcid=,sname='Alonso']{Enrique Alonso}
\affiliation{Fundación Ceo, Ciencia e Cultura, FC3/OAF, Campus Lagoas-Marcosende, s/n, 36310 Vigo, Spain}
\email{oaf.enrique@gmail.com}

\author[orcid=,sname='Amado']{Pedro J. Amado}
\affiliation{Instituto de Astrofísica de Andalucía (IAA-CSIC), Gta. de la Astronomía s/n, 18008 Granada, Spain}
\email{pedro.amado@iaa.csic.es}

\author[orcid=0000-0002-4746-0181,sname='Bacsturk']{\"Ozg\"ur Ba\c{s}t\"urk}
\affiliation{Department of Astronomy \& Space Sciences, Faculty of Science, Ankara University, TR-06100, Ankara, T\"urkiye}
\affiliation{Ankara University, Astronomy and Space Sciences Research and Application Center (Kreiken Observatory), Incek Blvd., TR-06837, Ahlatlıbel, Ankara, T\"urkiye}
\email{obasturk@ankara.edu.tr} 

\author[0000-0002-9003-484X]{David Charbonneau}
\affiliation{Center for Astrophysics \textbar\ Harvard \& Smithsonian, 60 Garden Street, Cambridge, MA 02138, USA}
\email{dcharbonneau@cfa.harvard.edu}

\author[orcid=0000-0002-5741-3047,sname='Ciardi']{David~R.~Ciardi}
\affiliation{NASA Exoplanet Science Institute, Caltech/IPAC, Pasadena, CA 91125, USA}
\email{ciardi@ipac.caltech.edu}

\author[orcid=0000-0002-2361-5812,sname='Clark']{Catherine A.~Clark}
\affiliation{NASA Exoplanet Science Institute, Caltech/IPAC, Pasadena, CA 91125, USA}
\email{clarkc@ipac.caltech.edu} 

\author[orcid=,sname='Fern\'andez-Rodr\'iguez']{Gareb Fern\'andez-Rodr\'iguez}
\affiliation{Instituto de Astrof\'isica de Canarias (IAC), E-38200 La Laguna, Tenerife, Spain}
\affiliation{Departamento de Astrof\'isica, Universidad de La Laguna (ULL), E-38206 La Laguna, Tenerife, Spain}
\email{gareb.fernandez@iac.es} 

\author[orcid=0000-0002-9436-2891,sname='Fukuda']{Izuru Fukuda}
\affiliation{Department of Multi-Disciplinary Sciences, Graduate School of Arts and Sciences, The University of Tokyo, 3-8-1 Komaba, Meguro, Tokyo 153-8902, Japan}
\email{izuru-fukuda@g.ecc.u-tokyo.ac.jp} 

\author[orcid=0000-0002-4909-5763,sname='Fukui']{Akihiko Fukui}
\affiliation{Komaba Institute for Science, The University of Tokyo, 3-8-1 Komaba, Meguro, Tokyo 153-8902, Japan}
\affiliation{Instituto de Astrof\'{i}sica de Canarias (IAC), 38205 La Laguna, Tenerife, Spain}
\email{afukui@g.ecc.u-tokyo.ac.jp} 

\author[orcid=0000-0001-9800-6248,sname='Furlan']{Elise Furlan}
\affiliation{NASA Exoplanet Science Institute, Caltech/IPAC, Pasadena, CA 91125, USA}
\email{furlan@ipac.caltech.edu} 

\author[orcid=0000-0002-4503-9705,sname='Gan']{Tianjun Gan}
\affiliation{Instituto de Astrof\'isica de Canarias (IAC), E-38200 La Laguna, Tenerife, Spain}
\affiliation{Departamento de Astrof\'isica, Universidad de La Laguna (ULL), E-38206 La Laguna, Tenerife, Spain}
\email{tianjungan@gmail.com} 

\author[0000-0003-1361-985X]{Juliana Garc\'ia-Mej\'ia}
\altaffiliation{51 Pegasi b Fellow, MIT Pappalardo Physics Fellow}
\affiliation{Kavli Institute for Astrophysics and Space Research, Massachusetts Institute of Technology, Cambridge, MA 02139, USA}
\affiliation{Center for Astrophysics \textbar\ Harvard \& Smithsonian, 60 Garden Street, Cambridge, MA 02138, USA}
\email{juliana.garcia-mejia@cfa.harvard.edu} 

\author[orcid=0009-0009-7996-8703,sname='García-Serra']{Andrés García-Serra}
\affiliation{Instituto de Astrof\'isica de Canarias (IAC), E-38200 La Laguna, Tenerife, Spain}
\affiliation{Departamento de Astrof\'isica, Universidad de La Laguna (ULL), E-38206 La Laguna, Tenerife, Spain}
\email{alu0101451923@ull.edu.es} 

\author[orcid=,sname='Geraldía-González']{Samuel Geraldía-González}
\affiliation{Instituto de Astrof\'isica de Canarias (IAC), E-38200 La Laguna, Tenerife, Spain}
\affiliation{Departamento de Astrof\'isica, Universidad de La Laguna (ULL), E-38206 La Laguna, Tenerife, Spain}
\email{samuel.geraldia.gonzalez@iac.es} 

\author[orcid=0000-0002-8965-3969,sname='Giacalone']{Steven Giacalone}
\affiliation{NSF Astronomy and Astrophysics Postdoctoral Fellow}
\affiliation{Department of Astronomy, California Institute of Technology, Pasadena, CA 91125, USA}
\email{giacalone@astro.caltech.edu} 

\author[orcid=,sname='González']{Hugo González}
\affiliation{Fundación Ceo, Ciencia e Cultura, FC3/OAF, Campus Lagoas-Marcosende, s/n, 36310 Vigo, Spain}
\email{hugogr321@gmail.com}

\author[orcid=0000-0001-8877-0242,sname='Hayashi']{Yuya Hayashi}
\affiliation{Department of Multi-Disciplinary Sciences, The University of Tokyo, 3-8-1 Komaba, Meguro, Tokyo 153-8902, Japan}
\email{yuya-hayashi@g.ecc.u-tokyo.ac.jp} 

\author[orcid=0000-0002-2532-2853,sname='Howell']{Steve~B.~Howell}
\affiliation{NASA Ames Research Center, Moffett Field, CA 94035, USA}
\email{steve.b.howell@nasa.gov)} 

\author[orcid=0000-0002-6480-3799,sname='Isogai']{Keisuke Isogai}
\affiliation{Okayama Observatory, Kyoto University, 3037-5 Honjo, Kamogatacho, Asakuchi, Okayama 719-0232, Japan}
\affiliation{Department of Multi-Disciplinary Sciences, Graduate School of Arts and Sciences, The University of Tokyo, 3-8-1 Komaba, Meguro, Tokyo 153-8902, Japan}
\email{isogai@kusastro.kyoto-u.ac.jp} 

\author[orcid=0000-0002-4197-7374,sname='Lacedelli']{Gaia Lacedelli}
\affiliation{Instituto de Astrof\'isica de Canarias (IAC), E-38200 La Laguna, Tenerife, Spain}
\affiliation{Departamento de Astrof\'isica, Universidad de La Laguna (ULL), E-38206 La Laguna, Tenerife, Spain}
\email{gaia.lacedelli@iac.es} 

\author[orcid=,sname='Libotte']{Florence Libotte}
\affiliation{Instituto de Astrof\'isica de Canarias (IAC), E-38200 La Laguna, Tenerife, Spain}
\affiliation{Sabadell Astronomical Society, Sabadell, Spain}
\email{florence.libotte.astro@gmail.com} 

\author[orcid=0000-0003-2527-1598,sname='Lund']{Michael B.~Lund}
\affiliation{NASA Exoplanet Science Institute, Caltech/IPAC, Pasadena, CA 91125, USA}
\email{mlund@ipac.caltech.edu}

\author[orcid=0009-0001-7943-0075,sname='Meni-Gallardo']{Pedro Meni-Gallardo}
\affiliation{Instituto de Astrof\'isica de Canarias (IAC), E-38200 La Laguna, Tenerife, Spain}
\affiliation{Departamento de Astrof\'isica, Universidad de La Laguna (ULL), E-38206 La Laguna, Tenerife, Spain}
\email{pepamanguer@gmail.com} 

\author[orcid=,sname='López-Puertas']{Manuel López-Puertas}
\affiliation{Instituto de Astrofísica de Andalucía (IAA-CSIC), Gta. de la Astronomía s/n, 18008 Granada, Spain}
\email{puertas@iaa.es}

\author[orcid=0000-0002-4671-2957,sname='Luque']{Rafael Luque}
\affiliation{Instituto de Astrofísica de Andalucía (IAA-CSIC), Gta. de la Astronomía s/n, 18008 Granada, Spain}
\email{rluque@iaa.es} 

\author[orcid=0000-0001-9087-1245,sname='Murgas']{Felipe Murgas}
\affiliation{Instituto de Astrof\'isica de Canarias (IAC), E-38200 La Laguna, Tenerife, Spain}
\affiliation{Departamento de Astrof\'isica, Universidad de La Laguna (ULL), E-38206 La Laguna, Tenerife, Spain}
\email{fmurgas@iac.es} 

\author[orcid=0009-0002-6086-2086,sname='Parlapani']{Zoe Parlapani}
\affiliation{Delft University of Technology}
\email{zoe.parlapani@gmail.com} 

\author[orcid=,sname='Parviainen']{Hannu Parviainen}
\affiliation{Instituto de Astrof\'isica de Canarias (IAC), E-38200 La Laguna, Tenerife, Spain}
\affiliation{Departamento de Astrof\'isica, Universidad de La Laguna (ULL), E-38206 La Laguna, Tenerife, Spain}
\email{hpparvi@gmail.com} 

\author[orcid=0000-0003-1337-723X,sname='Rockcliffe']{Keighley E. Rockcliffe}
\affiliation{University of Maryland Baltimore County, Baltimore, MD, USA}
\affiliation{NASA Goddard Space Flight Center, Greenbelt, MD, USA}
\email{keighley.e.rockcliffe@nasa.gov} 

\author[orcid=0000-0003-1713-3208,sname='Safonov']{Boris Safonov}
\affiliation{Sternberg Astronomical Institute Lomonosov Moscow State University.}
\email{safonov10@gmail.com}

\author[orcid=0000-0003-2693-279X,sname='Sánchez-Benavente']{Manuel Sánchez-Benavente}
\affiliation{Instituto de Astrof\'isica de Canarias (IAC), E-38200 La Laguna, Tenerife, Spain}
\affiliation{Departamento de Astrof\'isica, Universidad de La Laguna (ULL), E-38206 La Laguna, Tenerife, Spain}
\email{manuelsb93@gmail.com} 

\author[orcid=0000-0002-0516-7956,sname='Sánchez-López']{Alejandro Sánchez-López}
\affiliation{Instituto de Astrofísica de Andalucía (IAA-CSIC), Gta. de la Astronomía s/n, 18008 Granada, Spain}
\email{alexsl@iaa.es}

\author[orcid=0000-0002-1420-1837,sname='Softich']{Emma Softich}
\affiliation{Department of Astronomy \& Astrophysics, University of California, San Diego, La Jolla, CA, USA}
\email{esoftich@ucsd.edu}

\author[orcid=0000-0001-8227-1020,sname='Schwartz']{Richard P. Schwarz}
\affiliation{Center for Astrophysics | Harvard \& Smithsonian, 60 Garden Street, Cambridge, MA 02138, USA}
\email{rpschwarz@comcast.net}

\author[orcid=0000-0003-1038-9702,sname='Scott']{Nicholas~J.~Scott}
\affiliation{The CHARA Array of Georgia State University, Mount Wilson Observatory, Mt. Wilson, CA 91203, USA}
\email{nic.scott.9@gmail.com} 

\author[orcid=0000-0001-6424-5005,sname='Ulla Miguel']{Ana Ulla Miguel}
\affiliation{Applied Physics Department, Universidade de Vigo, Campus Lagoas-Marcosende, s/n, 36310 Vigo, Spain}
\affiliation{Centro de Investigación Mariña, Universidade de Vigo, CIM/GEOMA, 36310 Vigo, Spain}
\affiliation{Fundación Ceo, Ciencia e Cultura, FC3/OAF, Campus Lagoas-Marcosende, s/n, 36310 Vigo, Spain}
\email{ulla@uvigo.gal} 

\author[orcid=0000-0003-2192-5371,sname='van Eyken']{Julian C. van Eyken}
\affiliation{NASA Exoplanet Science Institute, Caltech/IPAC, Pasadena, CA 91125, USA}
\email{vaneyken@ipac.caltech.edu}

\author[orcid=,sname='Yalcinkaya']{Sel\c{c}uk Yal\c{c}{\i}nkaya}
\affiliation{Department of Astronomy \& Space Sciences, Faculty of Science, Ankara University, TR-06100, Ankara, T\"urkiye}
\affiliation{Ankara University, Astronomy and Space Sciences Research and Application Center (Kreiken Observatory), Incek Blvd., TR-06837, Ahlatlıbel, Ankara, T\"urkiye}
\affiliation{Astrobiology Research Unit, Universit\'e de Li\`ege, All\'ee du 6 Ao\^ut 19C, B-4000 Li\`ege, Belgium}
\email{selcuk_yalcinkaya@yahoo.com}

\author[orcid=,sname='Zambelli']{Roberto Zambelli}
\affiliation{American Association of Variable Star Observers, 185 Alewife Brook Parkway, Suite 410, Cambridge, MA 02138, USA}
\affiliation{Società Astronomica Lunae, Castelnuovo Magra, Italy}
\email{robertozambelli.rz@libero.it}

\begin{abstract}

We identify two planets in close-in orbits around the M-type star TOI-2079 as the first super-Earth-sized planets orbiting a multi-planet M dwarf with a well-determined age of $1$--$2$\,Gyr. The interior planet, TOI-2079\,b, has an orbital period of $1.49232359^{+0.00000099}_{-0.00000086}$\,d, and the outer planet, TOI-2079\,c, has an orbital period of $9.3174523\pm0.0000098$\,d. The host star characterization is based on spectroscopic observations, which classify it from an M1.5 to an M3.5 dwarf with an effective temperature of $3695\pm77$\,K. Further analysis determined TOI-2079's rotation period to be \Prot~d. Multiple ground-based follow-up observations, combined with a state-of-the-art validation procedure, allow us to rule out possible astrophysical false-positive scenarios and validate the planetary nature of both objects. We find that the inner planet has a radius of $1.320\pm0.050\,R_{\oplus}$ and $T_{\rm eq}\,=\,867\pm27$\,K, while the outer planet has $R_{\rm p}\,=\,1.671\pm0.051\,R_{\oplus}$ and $T_{\rm eq}\,=\,473\pm15$\,K.

\end{abstract}


\section{Introduction}
\label{sec:introduction}

Exoplanet discoveries in the past three decades have shown that there is a rich variety of planetary systems within our galaxy, exhibiting diverse physical properties and orbital configurations. Space-based transit surveys, particularly Kepler \citep{2010Sci...327..977B} and TESS \citep{2015JATIS...1a4003R}, have been fundamental to building this picture, with nearly $5000$ confirmed planets discovered through the transit method, corresponding to $\sim 75\%$ of the total. Demographic studies have revealed characteristics such as the radius gap, which is an occurrence drop for exoplanets with radii around $\rm 1.6~R_{\oplus}$, constituting a boundary between super-Earths and sub-Neptunes \citep{2017AJ....154..109F}. The most accepted explanations for this feature involve atmospheric mass loss, either photoevaporation, core-powered mass loss, or both, depending on the age of the system \citep{2024MNRAS.528.1615O, 2026arXiv260502766M}. Furthermore, planets discovered in an evolved stage might present a different bulk and atmospheric composition than at the time of their formation due to other processes, such as planet migration through the disk, mantle outgassing, and meteorite bombardment \citep{Baruteau2016, 2018ARA&A..56..175D, 2018A&A...614A..18D, owenLai2018, 2022ApJ...938L...3L}.

Stellar age is therefore a key parameter for understanding the evolution of planetary populations and their present-day demographics. Yet, only a small fraction of transiting planets orbit stars with well-constrained ages and spectral types \citep{2016RSOS....350571V, 2019ApJ...880L..17N}, limiting our ability to trace planetary evolution across time. While several $<$1\,Gyr exoplanets have been recently discovered (e.g., \citealp{Barber2024,Vach2025,Barber2026}), only a few planets have been discovered orbiting intermediate-aged, or adolescent, M dwarfs, the most abundant type of star in our Galaxy (e.g., TOI-700; \citealp{2020AJ....160..116G}). If we additionally consider only multi-planet systems, there are, to date, only six known systems, none of which host super-Earth-sized planets. With orbital periods of $\sim100$\,days and sizes of $\lesssim1.8\,R_{\oplus}$, super-Earths are among the most common classes of planets in our Galaxy, offering unique insights into planet formation \citep{2012ApJS..201...15H, 2013ApJ...766...81F, 2013PNAS..11019273P, 2015ApJ...814..130M, 2023AJ....165..262Z, 2025ApJ...979...79R}.

In this Letter, we present the discovery of TOI-2079, a system hosting at least two planets: the close-in, warm, super-Earth-sized planet TOI-2079\,b and the more distant, larger super-Earth TOI-2079\,c. These planets constitute the first multi-super-Earth system around an adolescent M dwarf. The host star is a bright ($K\,=\,8.6$\,mag), nearby ($37.1$\,pc), M1.5 dwarf star with an estimated age of 1-2 Gyr that is observable from the northern hemisphere (see Appendix\,\ref{sec:observations_p_appendix}). We describe the photometric observations of the system in Section\,\ref{sec:observations} and the characterization of the host star in Section\,\ref{sec:star}. We validate the planetary system in Section\,\ref{sec:validation}, presenting the light curves analysis and modeling of the planetary parameters. Section\,\ref{sec:discussion_conclusion} concludes our study, highlighting the uniqueness of this system in the context of other benchmark planets. Appendices\,\ref{sec:observations_p_appendix}, \ref{sec:observations_s_appendix}, \ref{sec:hr_imaging_appendix}, \ref{sec:transit_analysis_appendix}, \ref{sec:validation_appendix}, and \ref{sec:prospects_appendix} respectively include further details on the optical observations, spectroscopic observations, high-resolution imaging, transit analysis, system validation, and prospects for mass determination and atmospheric characterization.

\section{Observations}
\label{sec:observations}

TOI-2079 was observed in $5$ TESS sectors ($15$, $22$, $41$, $48$, $75$) at $120$\,s cadence. We used ground-based follow-up photometry from several facilities in the context of the TESS Follow-up Observing Program (TFOP) Sub-Group 1 (SG1) for Seeing-Limited Photometry. The main goals of these observations were to confirm the signals corresponding to TOI-2079.01 and TOI-2079.02 in the target star at transit times, obtain light curves with higher precision than TESS data, and evaluate the transit depths in different bands to assess the chromatic dependence. A complete description of the photometry together with the TESS image around the position of TOI-2079 in the selected sectors can be found in Appendix\,\ref{sec:observations_p_appendix}.

In order to properly characterize the stellar parameters, we observed TOI-2079 with the SpeX spectrograph \citep{Rayner2003} on the 3.2\,m NASA Infrared Telescope Facility (IRTF) on May 01, 2026 (UT) in clear conditions and seeing of $1\farcs2$, and with the Kast double spectrograph \citep{kastspectrograph} on the 3\,m Shane telescope at Lick Observatory at the end of the night on 2025 January 19 (UT) also in clear conditions with 1$\arcsec$ seeing (see Appendix\,\ref{sec:observations_s_appendix}). We have also analyzed high-resolution imaging observations of TOI-2079 as described in Appendix\,\ref{sec:hr_imaging_appendix}. These observations are used in order to asses any possible contamination of a companion on the derived planetary parameters.

\section{The host star}
\label{sec:star} 

In this section, we present the spectroscopic analysis of the observed spectra and the derived stellar parameters, including constraints on the stellar age.

\subsection{Spectroscopic analysis}
\label{subsec:spec_analysis}

We compared the SXD spectrum to single-star standards from the IRTF Spectral Library \citep{Cushing2005, Rayner2009} using the SpeX Prism Library Analysis Toolkit \citep[SPLAT, ][]{splat}.
The best match is the M1.5\,V standard HD\,36395, and we therefore adopt a near-infrared spectral type of M1.5$\pm$0.5 for TOI-2079,
consistent with prior infrared classifications \citep{2015ApJS..220...16T,2020ApJ...892...31B}.
Using the $K$-band Na\,\textsc{i} and Ca\,\textsc{i} features and the H$_2$O--K2 index \citep{Rojas-Ayala2012}, we estimate $\mathrm{[Fe/H]} = -0.01 \pm 0.11$ from the \citet{Mann2013} relation,
formally consistent with APOGEE DR17 metallicity $\mathrm{[Fe/H]} = -0.092 \pm 0.008$ \citep{accetta2022seventeenth}.

The reduced Kast optical spectrum is compared to its best-fit M2 dwarf SDSS spectral template from 
\citet{2007AJ....133..531B}.
This template classification is consistent with the infrared classification, while index-based classifications from 
\citet{1995AJ....110.1838R,1997AJ....113..806G,1999AJ....118.2466M,2003AJ....125.1598L}; and \citet{2007MNRAS.381.1067R} yield somewhat later types of M2.5--M3.5. Despite the poor S/N, we detected weak 6563~{\AA} H$\alpha$ emission with an equivalent width of $-$0.7$\pm$0.3~{\AA}, corresponding to moderate emission of $\log{L_{H\alpha}/L_{bol}} = -4.43\pm0.19$ based on the $\chi$-factor correction of \citet{2014ApJ...795..161D}. 
The 4681~{\AA} H$\beta$ line is not detected with sufficient S/N. We measured a zeta metallicity index of $\zeta$=1.15$\pm$0.02  \citep{2013AJ....145..102L}, corresponding to near solar metallicity ([Fe/H] = $+0.19\pm0.20$; \citealt{2013AJ....145...52M}), again consistent with infrared measurements reported here and in prior studies \citep{2020ApJ...892...31B,2025ApJ...984...58H}.

Spectra from both SXD and Kast can be found in Appendix\,\ref{sec:observations_s_appendix}. Updated stellar parameters derived from empirical relations of \cite{Mann2015, Mann2019} are shown in Table\,\ref{tab:parameters}.

\subsection{Stellar age}
\label{subsec:stellar_age}

Several independent indicators constrain the age of TOI-2079. The Kast spectrum shows hydrogen emission, indicating an activity age $\lesssim$1$-$2\,Gyr for this spectral type \citep{West2008}. This is consistent with the ASAS-SN light curve \citep[RMS $=$ 0.015 mag over $\sim$ 5 yr]{Shappee2014, Kochanek2017}, which shows no large, long-duration flaring events and disfavors a strongly active, young M dwarf. GALEX NUV photometry and FUV$-$NUV color \citep{Martin2005}, compared to the age$-$activity relations of \citet{MilesShkolnik2017} and \citet{SchneiderShkolnik2018}, indicate an age not younger than $\sim$1\,Gyr, though the FUV detection is only 2$\sigma$ and should be treated as a lower limit on the true FUV$-$NUV color. The stellar Ly$\alpha$ emission was also constrained by the STEL$\alpha$ program (Loyd et al. in preparation). The reconstructed Ly$\alpha$ luminosity, $\mathrm{log}(L_\mathrm{Ly\alpha}/L_{\rm bol}) = -4.38^{+0.472}_{-0.421}$ (Rockcliffe et al., in preparation), is atypical for stars younger than $\sim$1 Gyr when compared to the tracks of \citet{Peacock2025}, though the tracks and measurement uncertainties do not exclude ages up to several Gyr (see Appendix\,\ref{subsec:spectroscopy_appendix_hst}). BANYAN $\Sigma$ \citep{2018ApJ...856...23G} finds no association with any known young moving group \citep{Gagne2026, Way2026}, ruling out ages $\lesssim$100$-$300 Myr but not further constraining the field-age range.

{\it Tierras} photometry indicates a rotation period of \Prot~d (Fig.~\ref{fig:tierras}; see Appendix~\ref{subsec:Tierras}), which corresponds well with the 650~Myr Praesepe $P_\mathrm{rot}-T_\mathrm{eff}$ sequence \citep{Rampalli2021, Bouma2023}. Early-to-mid M dwarfs are expected to have already converged onto this slow-rotator sequence by Praesepe's age \citep{Popinchalk2021}, so TOI-2079 position there is expected. This would suggest a young/intermediate age, but it should not be read as a precise constraint, for two reasons. First, at $T_{\rm eff}\approx$ 3695 K, TOI-2079 is cooler than the 3800--6200 K range which is currently used in gyrochronology calibrations for validation \citep{Bouma2023}. That is, there are not enough benchmark M-dwarf rotators at intermediate-to-old ages this cool to calibrate against. Second, stars with estimated masses close to TOI-2079 are known to stall: after reaching the slow-rotator sequence, they can sit at a nearly constant period for a long duration before spinning down further. This has been shown for K and early M dwarfs down to $\sim$0.55 M$_\sun$ \citep{Curtis2020}, and more recently for the CARMENES M-dwarf sample directly \citep{Shan2024}.

\begin{figure}[h!]
    \centering
    \includegraphics[width=\columnwidth]{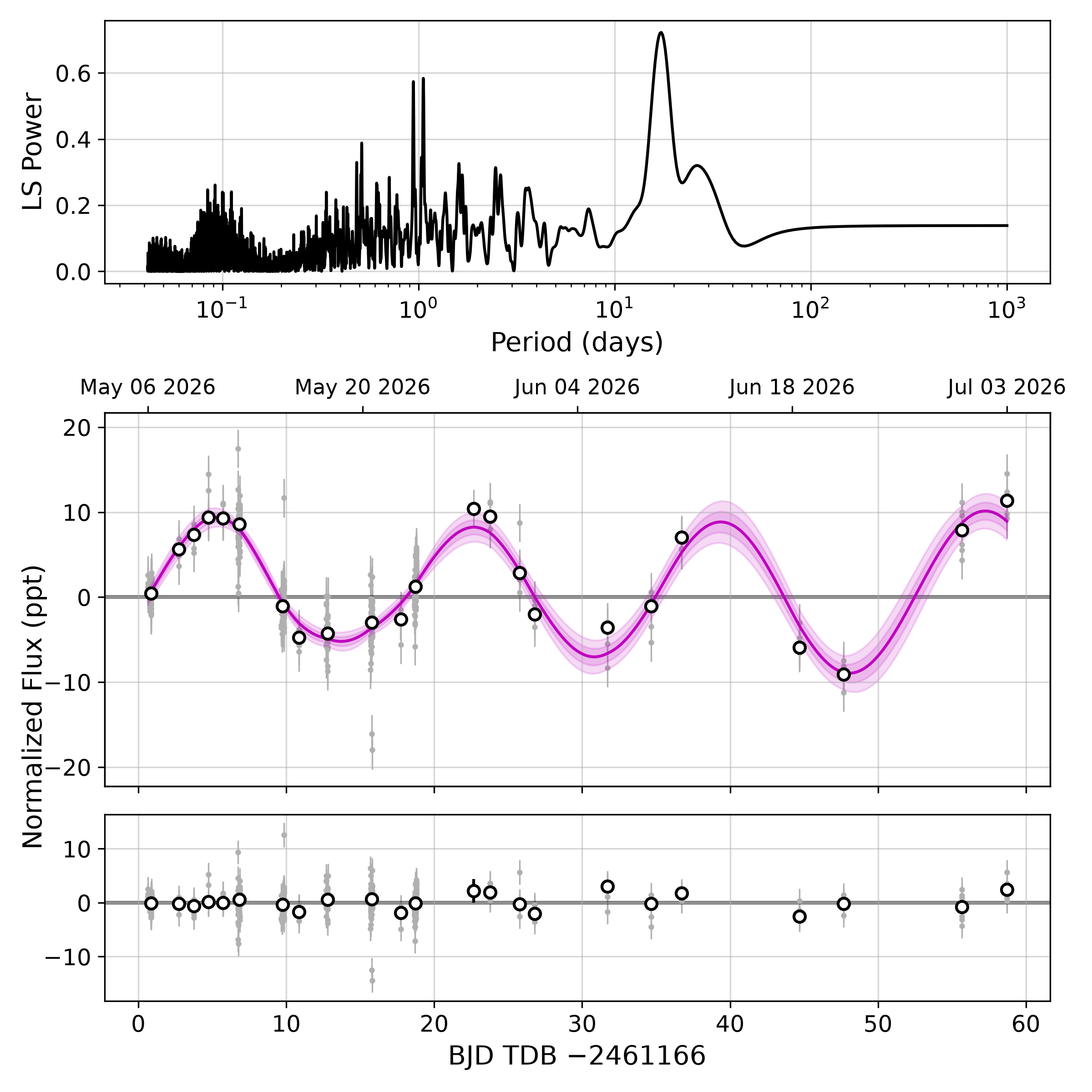}
    \caption{Top: The Lomb-Scargle periodogram of the \textit{Tierras} light curve, the highest peak of which is at 17.3~days. Middle: The \textit{Tierras} light curve. Individual exposures, taken with a 60-s exposure time, are shown in dots with error bars, while the median of each night is indicated with a white circle with black outline. The maximum \textit{a posteriori} (MAP) quasi-periodic Gaussian process model is shown in magenta, along with its 1- and 2-$\sigma$ uncertainty contours (see Appendix~\ref{subsec:Tierras}). Bottom: The light curve residuals from the MAP model in the middle panel.}
    \label{fig:tierras}
\end{figure}

Taken together, these indicators suggest that TOI-2079 is an adolescent star between 500~Myr and 2~Gyr in age. The UV/Ly$\alpha$ activity indicators set a lower bound of $\sim$1 Gyr and the detection of hydrogen emission and the lack of ASAS-SN flaring set an upper bound of $\sim$1$-$2 Gyr. The measured rotation period is consistent with the younger end of this range, though we should treat this estimation with caution for the detailed reasons above. On balance, the evidence favors an age of 1$-$2 Gyr, though a younger age within the broader 0.5$-$2 Gyr range cannot be excluded given the uncertainties inherent to each method.

\section{TOI-2079 system validation}
\label{sec:validation}

We analyzed all available 2-minute cadence \textit{TESS} PDCSAP light curves of TOI-2079 using the \texttt{SHERLOCK} pipeline \citep{devorapajares2024, 2026MNRAS.549ag183M, 2026MNRAS.548ag713P}. Using all available sectors ($15$, $22$, $41$, $48$, and $75$), \texttt{SHERLOCK} recovered the known planets TOI-2079.01 and TOI-2079.02 at high signal-to-noise ratio. No additional statistically significant periodic transit signals were identified after masking these planets or using \texttt{MORIARTY} (the single-transit detection module within the \texttt{SHERLOCK} framework), indicating no evidence for further transiting companions within the sensitivity limits of the available \textit{TESS} photometry (see Appendix\,\ref{sec:transit_analysis_appendix}).

\begin{figure}
    \centering
    \includegraphics[width=\linewidth]{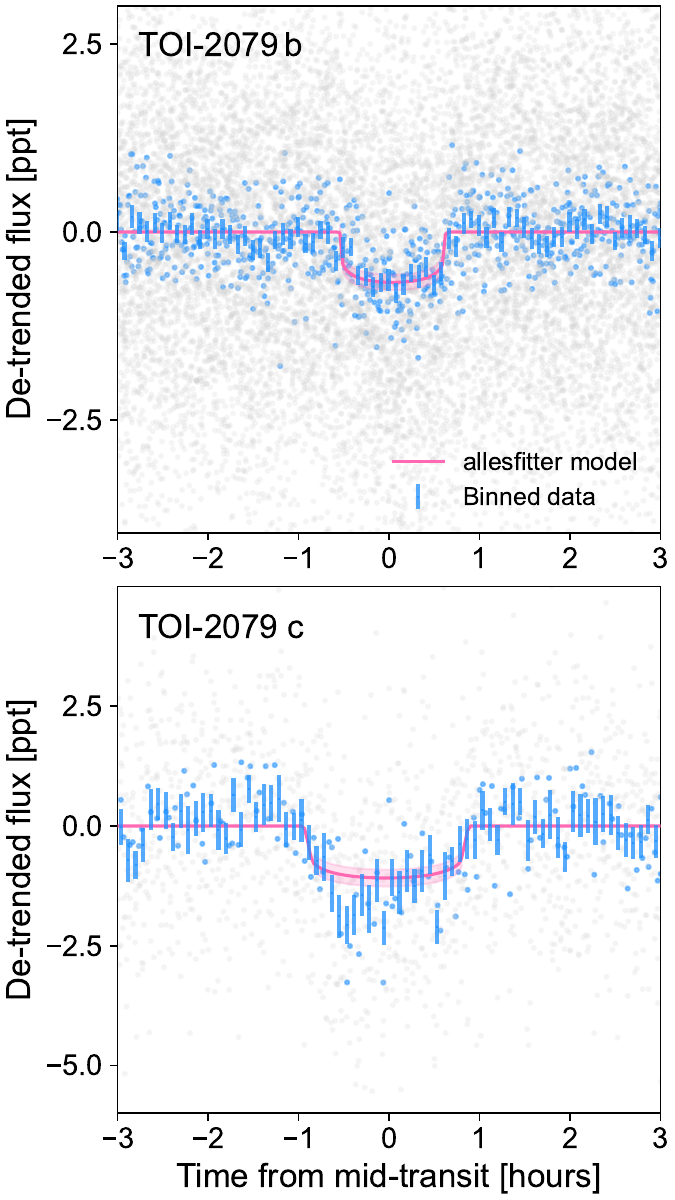}
    \caption{Phase folded light curves from the TESS sectors for planets TOI-2079.01 (top panel) and TOI-2079.02 (bottom panel), together with the transit models (pink line) showing the $1\sigma$ range (pink region) in models during transit. Observed data is binned in steps of half a minute (blue dots) and $5$\,minutes (blue lines) in phase for TOI-2079\,b, and steps of $2$\,minutes and $5$ minutes in phase for TOI-2079\,c.}
    \label{fig:global_model}
\end{figure}

The resulting light curves from our combined model, which includes all good-quality light curves, are shown in Fig.\,\ref{fig:global_model} (see also Appendix\,\ref{sec:observations_s_appendix}). The model and derived physical parameters are reported in Table\,\ref{tab:parameters}. The planetary masses and stellar radial-velocity semi-amplitudes are predicted using \texttt{SPRIGHT} \citep{Parviainen2024}.

\begin{table}
    \centering
    \caption{TOI-2079 system parameters}
    \footnotesize
    \begin{tabular}{c c c}
    
    \hline\hline
        \noalign{\smallskip}
        Parameter & Value & \\
        \noalign{\bigskip}
        
        TOI-2079 & TIC 224298134 &  \\
        \hline
        
        \noalign{\smallskip}
        \multicolumn{3}{c}{\textit{Stellar atmospheric parameters}} \\
        \noalign{\smallskip}
        
         $T_{\rm eff}\,$(K) & $3695\pm77$ & \\
         $\log{g}\,$(dex) & $4.753\pm0.026$ & \\
         $[\mathrm{Fe}/\mathrm{H}]\,$(dex) & $-0.01\pm0.11$ & \\
         Type & M1.5$\pm$0.5 & \\
         
         \noalign{\smallskip}
         \multicolumn{3}{c}{\textit{Derived stellar parameters}} \\
         \noalign{\smallskip}
         
         $R_\star\,(R_{\odot})$ & $0.4792\pm 0.0134$ & \\
         $M_\star\,(M_{\odot})$ & $0.4747\pm0.0103$ & \\
         $\rho_\star\,(\rho_{\odot})$ & $4.32 \pm 1.15$ & \\
         \noalign{\bigskip}
         
         & TOI-2079\,b & TOI-2079\,c \\
        \hline
        
        \noalign{\smallskip}
        \multicolumn{3}{c}{\textit{Transit model parameters}} \\
        \noalign{\smallskip}
        
        $R_{\rm p}/R_{\star}$ & $0.02526\pm0.00065$ & $0.03206^{+0.00027}_{-0.00048}$ \\
        $b$ & $0.409^{+0.051}_{-0.060}$ & $0.688^{+0.030}_{-0.035}$ \\
        $P_{\rm orb}$\,(d) & $1.49232359^{+0.00000099}_{-0.00000086}$ & $9.3174523\pm0.0000098$ \\
        $T_0-2450000$ & $9867.0215^{+0.0005}_{-0.0004}$ & $9866.5991^{+0.0009}_{-0.0010}$ \\
        (BJD) & \\
        
        \noalign{\smallskip}
        \multicolumn{3}{c}{\textit{Derived transit parameters}} \\
        \noalign{\smallskip}

        $i$\,(deg) & $87.42^{+0.41}_{-0.38}$ & $88.71^{+0.09}_{-0.08}$ \\
        $T_{\rm 14}$\,(h) & $1.183\pm0.017$ & $1.586^{+0.061}_{-0.054}$ \\
        $\rho_{\star}$\,$(\rho_{\odot})$ & $4.52^{+0.31}_{-0.28}$ & $4.40^{+0.34}_{-0.31}$ \\

        \noalign{\smallskip}
        \multicolumn{3}{c}{\textit{Derived planet parameters}} \\
        \noalign{\smallskip}
  
        $R_{\rm p}$\,($R_{\oplus}$) & $1.320\pm0.050$ & $1.671\pm0.051$ \\
        $a$\,(au) & $0.02025\pm0.00073$ & $0.0680\pm0.0026$ \\
        $T_{\rm eq}$\,(K) & $867\pm27$
        & $473\pm15$ \\

        \noalign{\smallskip}
        \multicolumn{3}{c}{\textit{Empirical planet parameters}} \\
        \noalign{\smallskip}

        Predicted $M_{\rm p}$\,($M_{\oplus}$) & $2.4\pm0.4$ & $5.0^{+1.3}_{-2.4}$ \\
        Predicted $K_{\star}$\,(m/s) & $2.2\pm0.4$ & $2.5^{+0.7}_{-1.1}$ \\
        TSM & $9\pm2$ & $31^{+25}_{-8}$ \\
        ESM & $5.5\pm0.5$ & $1.5^{+0.2}_{-0.1}$ \\
        
        \noalign{\smallskip}
        \hline
        
    \end{tabular}
    \label{tab:parameters}
\end{table}

The analysis of the ground-based observations for TOI-2079.01 and TOI-2079.02 shows consistent transit depths across the different bands, indicating no significant wavelength dependence (see Appendix\,\ref{sec:validation_appendix}). Therefore, we proceed with the standard levels of validation for the planetary nature of the TOI-2079 system.

We performed a statistical validation analysis using the \texttt{TRICERATOPS} package \citep[see, e.g.,][]{2021AJ....161...24G, 2022AJ....163...99G}. For TOI-2079.01, we found FPP\,=\,$0.00012\pm0.00005$. For TOI-2079.02, we found FPP\,=\,$0.0046\pm0.0005$. For both cases, NFPP values are below 0.001. As an independent complement, we analyzed both candidates with \texttt{WATSON-Net} \citep{devorapajares2025}, obtaining calibrated planetary probabilities of $0.994\pm0.011$ and $0.832\pm0.292$ for planets b and c, respectively, both of which fall within the \textit{Likely Planet} regime defined by \texttt{WATSON-Net}. Hence, under these circumstances, both candidates are placed in the statistically validated planet category. Hereafter, we refer to TOI-2079.01 and TOI-2079.02 as TOI-2079\,b and TOI-2079\,c, respectively (see Appendix\,\ref{sec:validation_appendix}).

\section{Discussion and conclusions}
\label{sec:discussion_conclusion}

\begin{figure}
    \centering
    \includegraphics[width=\linewidth]{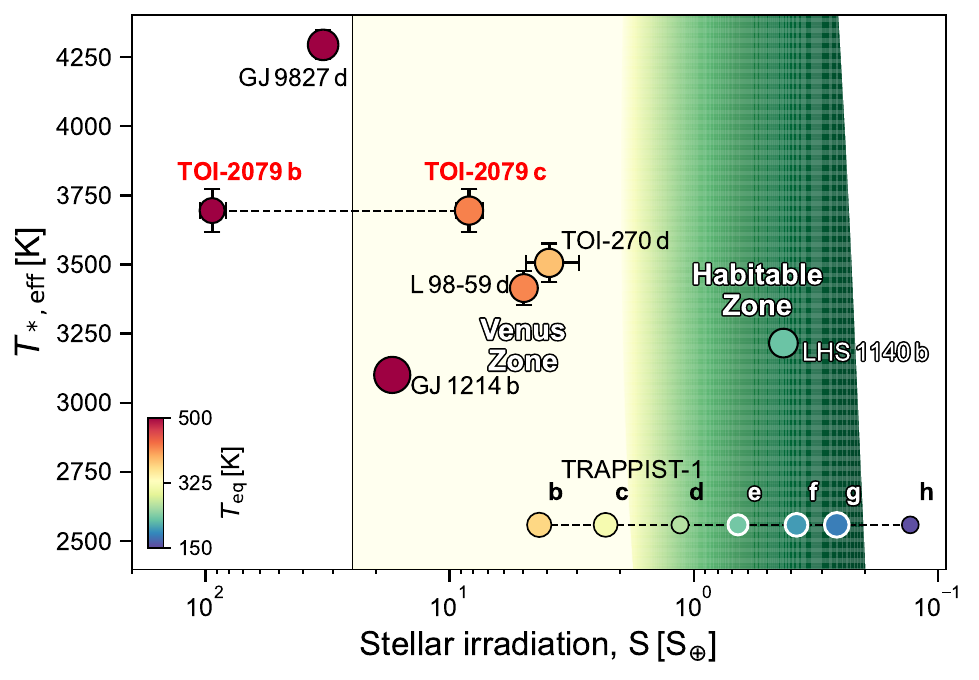}
    \caption{Stellar irradiation comparison between TOI-2079 planetary system, the sub-Neptune sample defined by \protect\cite{2025PNAS..12216194M}, and TRAPPIST-1. Planets are colored as a function of their $T_{\rm eq}$. We also present the gradual transition from the Venus Zone (VZ) (yellow shade; \protect\citealt{2014ApJ...794L...5K}) and the Habitable Zone (HZ) (green shade; \protect\citealt{2013ApJ...765..131K}).}
    \label{fig:vz_hz}
\end{figure}

\begin{figure}
    \centering
    \includegraphics[width=\linewidth]{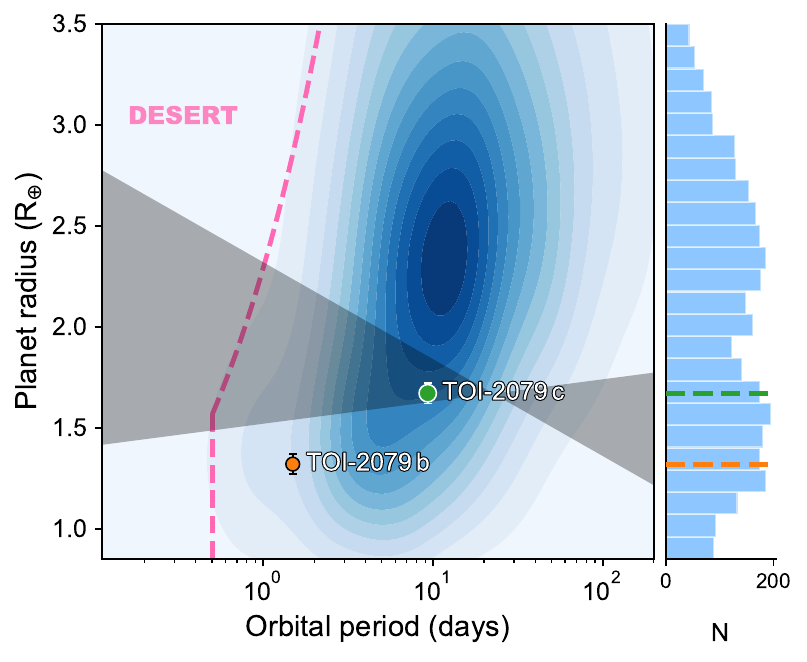}
    \caption{Location of the validated new planets (TOI-2079\,b in orange and TOI-2079\,c in green) in the period–radius diagram. The different contours and the corresponding color map indicate planet occurrence, with bluer regions being more populated and whiter regions less populated. The pink dashed line marks the boundary of the Neptune desert, as defined by \cite{2024A&A...689A.250C}. The black-shaded region shows the Radius Valley of sub-Neptune exoplanets around low-mass stars measured by \cite{2020AJ....159..211C}, and the Radius Valley for exoplanets around Sun-like stars measured by \cite{2019ApJ...875...29M} scaled to low-mass stars by \cite{2020AJ....159..211C}. The right panel shows the position of both planets in an occurrence rate histogram as a function of the planetary size.}
    \label{fig:neptune_desert}
\end{figure}

TOI-2079\,c, is placed in a particularly interesting region, the Radius Valley. Gyrochronological ages of FGK hosts indicate that the Radius Valley is in place within the first ~2 Gyr, with no significant change in its depth or width at later ages \citep{greco2026trackingevolutionsubneptunesii}. Planets in this size range lie at the boundary between the Earth-sized and mini-Neptune populations, and their ability to retain a gaseous envelope remains uncertain \citep{2014ApJ...783L...6W, 2015ApJ...801...41R}. Moreover, TOI-2079\,c, receives an incident flux compatible with the empirical Venus Zone \citep{2014ApJ...794L...5K}, making it a potential Venus analogue (see Fig.\,\ref{fig:vz_hz}). TOI-2079\,b, on the other hand, falls below this size gap, providing an interesting point of comparison within the system, with two planets orbiting the same star yet apparently following different evolutionary pathways. It also falls outside the current Neptune Desert definition proposed by \citep{2024A&A...689A.250C}. The classical formulation of the Neptune Desert by \cite{2016A&A...589A..75M} was more inclusive; however, it is now considered somewhat outdated, as several subsequent studies have refined the desert boundaries, restricting them to the regions that are genuinely depleted of planets \citep{2023A&A...677A..12D, 2023A&A...671A.132S, 2024A&A...692A.162M, 2024A&A...690A..62P}. The positions of TOI-2079\,b and c with respect to the Radius Valley and Neptune Desert are displayed in Fig.\,\ref{fig:neptune_desert}.

The population of multi-planet M dwarfs with well-constrained ages between $1$ and $2$\,Gyr and well-characterized spectral types is small. In fact, only six such systems are listed in the NASA Exoplanet Archive\footnote{\url{https://exoplanetarchive.ipac.caltech.edu}}: Kepler-52 \citep{2013MNRAS.428.1077S}, Kepler-125 \citep{2014ApJ...784...45R}, TOI-1470 \citep{2023A&A...675A.177G}, Kepler-94 \citep{2014ApJS..210...20M}, Kepler-210 \citep{2014ApJ...784...45R}, and Kepler-28 \citep{2012MNRAS.421.2342S}. With Kepler-125\,c being the smallest planet in this sample, with $R_{\rm p}=0.74\pm0.05\,R_{\oplus}$, and the closest in size to TOI-2079\,b, while Kepler-28\,c is the closest in size to TOI-2079\,c, with a radius of $1.857\pm0.042\,R{_\oplus}$, TOI-2079 remains the only known multi-planet system of this kind hosting two super-Earth-sized planets. Moreover, it shares similarities with the benchmark sub-Neptunes for JWST observations described in \cite{2025PNAS..12216194M}, particularly in terms of stellar irradiation and effective temperature (see Fig.\,\ref{fig:vz_hz}).

These characteristics make the TOI-2079 system an attractive laboratory for studying the atmospheric evolution of small planets. Our mass-precision estimates for MAROON-X indicate that a campaign of $\sim55$\,h of on-source time would yield mass precisions below 20\% for both targets (see Appendix\,\ref{subsec:prospects_appendix_mass}). We assess the prospects for JWST atmospheric and surface characterization, as well as for ground-based, high-resolution follow-up observations of possible He\,{\sc i} 10830\,\AA, in Appendix~F.

\begin{acknowledgments}

The TESS sectors presented in this article were obtained from the Mikulski Archive for Space Telescopes (MAST) at the Space Telescope Science Institute. The specific observations analyzed can be accessed via \dataset[doi: 10.17909/zdqy-gr62]{https://doi.org/10.17909/zdqy-gr62}.

All IAA-CSIC authors acknowledge financial support from the Severo Ochoa grant CEX2021-001131-S. APT acknowledges the financial support of the FPI-Severo Ochoa contract with reference PRE2022-104692 that is part of CEX2021-001131-S. GM acknowledges financial support from the Ramón y Cajal grant RYC2022-037854-I funded by MICIU/AEI/1144 10.13039/501100011033 and FSE+.

The work of B.S. was conducted under the state assignment of Lomonosov Moscow State University.

DRC acknowledges partial support from NASA Grant 18-2XRP18\_2-0007. This research has made use of the Exoplanet Follow-up Observation Program (ExoFOP; DOI: 10.26134/ExoFOP5) website, which is operated by the California Institute of Technology, under contract with the National Aeronautics and Space Administration under the Exoplanet Exploration Program. Based on observations obtained at the Hale Telescope, Palomar Observatory, as part of a collaborative agreement between the Caltech Optical Observatories and the Jet Propulsion Laboratory operated by Caltech for NASA. Some of the data presented herein were obtained at Keck Observatory, which is a private 501(c)3 non-profit organization operated as a scientific partnership among the California Institute of Technology, the University of California, and the National Aeronautics and Space Administration. The Observatory was made possible by the generous financial support of the W. M. Keck Foundation. The authors wish to recognize and acknowledge the very significant cultural role and reverence that the summit of Maunakea has always had within the Native Hawaiian community. We are most fortunate to have the opportunity to conduct observations from this mountain.

This material is based upon work supported by the National Aeronautics and Space Administration under Agreement No.\ 80NSSC21K0593 for the program ``Alien Earths.''
The results reported herein benefited from collaborations and/or information exchange within NASA's Nexus for Exoplanet System Science (NExSS) research coordination network sponsored by NASA's Science Mission Directorate.
This material is based upon work supported by the European Research Council (ERC) Synergy Grant under the European Union’s Horizon 2020 research and innovation program (grant No.\ 101118581---project REVEAL).

F. M. acknowledges the financial support from the Agencia Estatal de Investigaci\'{o}n del Ministerio de Ciencia, Innovaci\'{o}n y Universidades (MCIU/AEI) through grant PID2023-152906NA-I00.

This work is part of grant CEX2025-001609-S, awarded to the Instituto de Astrof\'{i}sica de Canarias under the Severo Ochoa Centre of Excellence program and funded by MICIU/AEI/10.13039/501100011033.

AUM acknowledges support by Autonomous Government Xunta de Galicia through grant ED431B 2026/45.

R.L. is funded by the European Union (ERC, THIRSTEE, 101164189) and acknowledges financial support from the Severo Ochoa grant CEX2021-001131-S funded by MCIN/AEI/10.13039/501100011033. Views and opinions expressed are however those of the author(s) only and do not necessarily reflect those of the European Union or the European Research Council. Neither the European Union nor the granting authority can be held responsible for them.

B.R-A acknowledges funding support from the ANID Basal project FB210003.

This work makes use of observations from the LCOGT network. Part of the LCOGT telescope time was granted by NOIRLab through the Mid-Scale Innovations Program (MSIP). MSIP is funded by NSF.

Funding for the TESS mission is provided by NASA's Science Mission Directorate. KAC acknowledges support from the TESS mission via subaward s3449 from MIT and NASA grants 80NSSC24K1889 and 80NSSC26K0081.

In this study, the observational data obtained within the scope of project numbered 22BT100-1958 and 25ATUG100-3012 carried out using the TUG100 telescope at the TUG (T\"UBİTAK National Observatory, Antalya) site of the T\"urkiye National Observatories has been utilised, and we express our gratitude for the invaluable support provided by the T\"urkiye National Observatories, the observation team, and all its staff. We gratefully acknowledge the support of the Scientific and Technological Research Council of T{\"u}rkiye (TÜBİTAK) with project 126F041.

The \textit{Tierras} Observatory is supported by the National Science Foundation under Award No.\ AST-2308043.
\end{acknowledgments}

Visiting Astronomer at the Infrared Telescope Facility, which is operated by the University of Hawaii under contract 80HQTR24DA010 with the National Aeronautics and Space Administration.

Some of the observations in the paper made use of the High-Resolution Imaging instrument ‘Alopeke. ‘Alopeke was funded by the NASA Exoplanet Exploration Program and built at the NASA Ames Research Center by Steve B. Howell, Nic Scott, Elliott P. Horch, and Emmett Quigley. ‘Alopeke was mounted on the Gemini North telescope of the international Gemini Observatory, a program of NSF’s OIR Lab, which is managed by the Association of Universities for Research in Astronomy (AURA) under a cooperative agreement with the National Science Foundation. on behalf of the Gemini partnership: the National Science Foundation (United States), National Research Council (Canada), Agencia Nacional de Investigación y Desarrollo (Chile), Ministerio de Ciencia, Tecnología e Innovación (Argentina), Ministério da Ciência, Tecnologia, Inovações e Comunicações (Brazil), and Korea Astronomy and Space Science Institute (Republic of Korea).

This article is based on observations made with the MuSCAT2 instrument, developed by ABC, at Telescopio Carlos Sánchez operated on the island of Tenerife by the IAC in the Spanish Observatorio del Teide. This work is partly supported by JSPS KAKENHI Grant Numbers JP18H01265 and JP18H05439, and JST PRESTO Grant Number JPMJPR1775. We acknowledge financial support from the Agencia Estatal de Investigaci\'on of the Ministerio de Ciencia e Innovaci\'on MCIN/AEI/10.13039/501100011033 and the ERDF “A way of making Europe” through projects PID2021-125627OB-C32 and PID2024-158486OB-C32. This work is supported by the ERC Grant (ERC Advanced Grant SPEAR, GA 101200674).  Funded by the European Union. Views and opinions expressed are however those of the author(s) only and do not necessarily reflect those of the European Union or the  European Research Council Executive Agency (ERCEA). Neither the European Union nor the granting authority can be held responsible for them.

This research has made use of the NASA Exoplanet Archive, which is operated by the California Institute of Technology, under contract with the National Aeronautics and Space Administration under the Exoplanet Exploration Program.

\appendix

\section{Photometric observations}
\label{sec:observations_p_appendix}

\begin{figure*}
    \centering
    \includegraphics[width=0.19\linewidth]{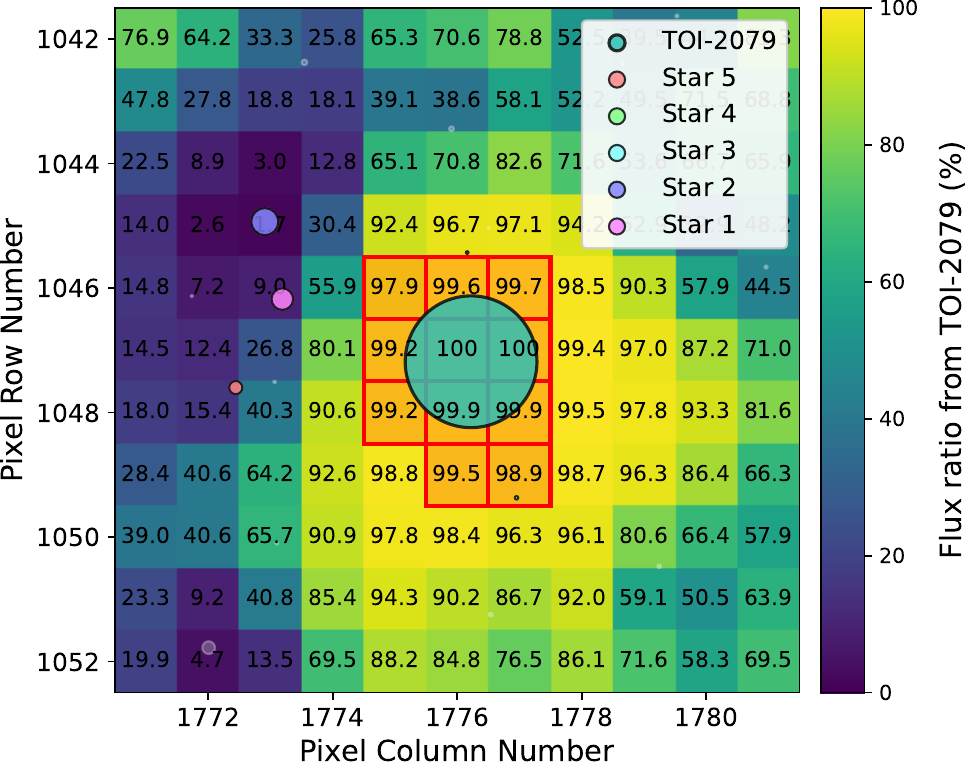}
    \includegraphics[width=0.19\linewidth]{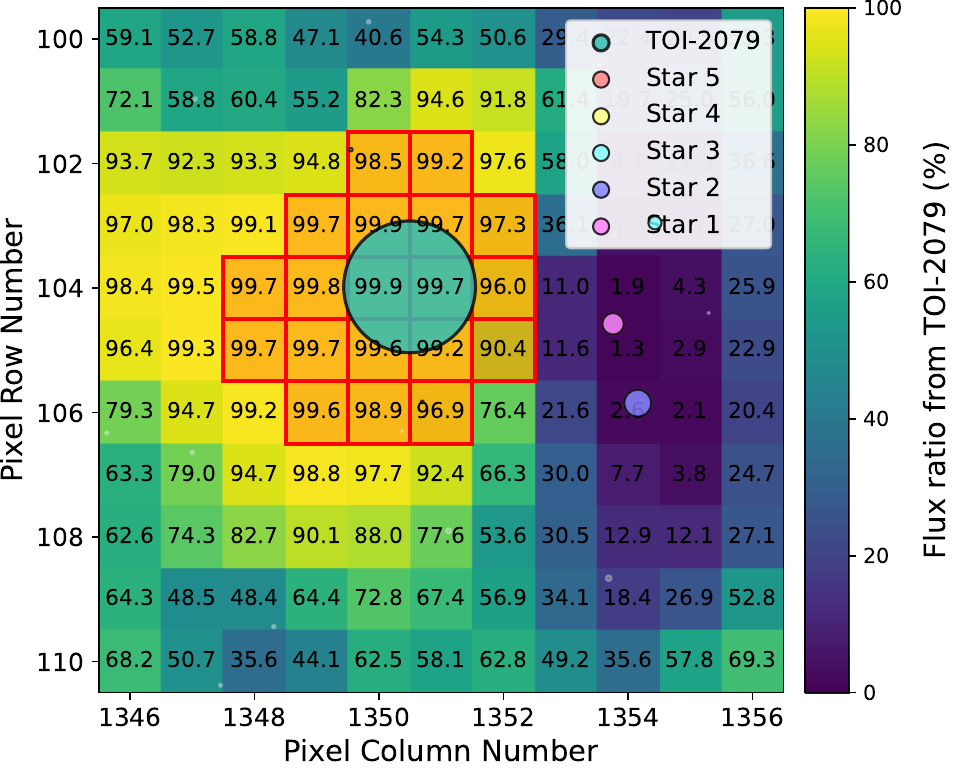}
    \includegraphics[width=0.19\linewidth]{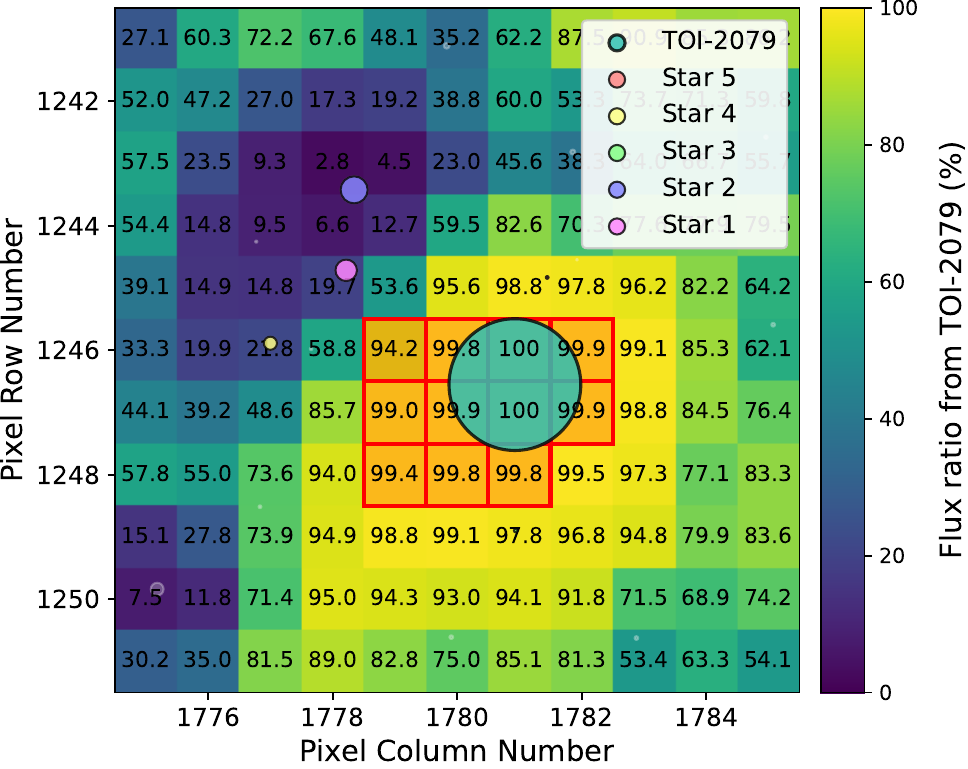}
    \includegraphics[width=0.19\linewidth]{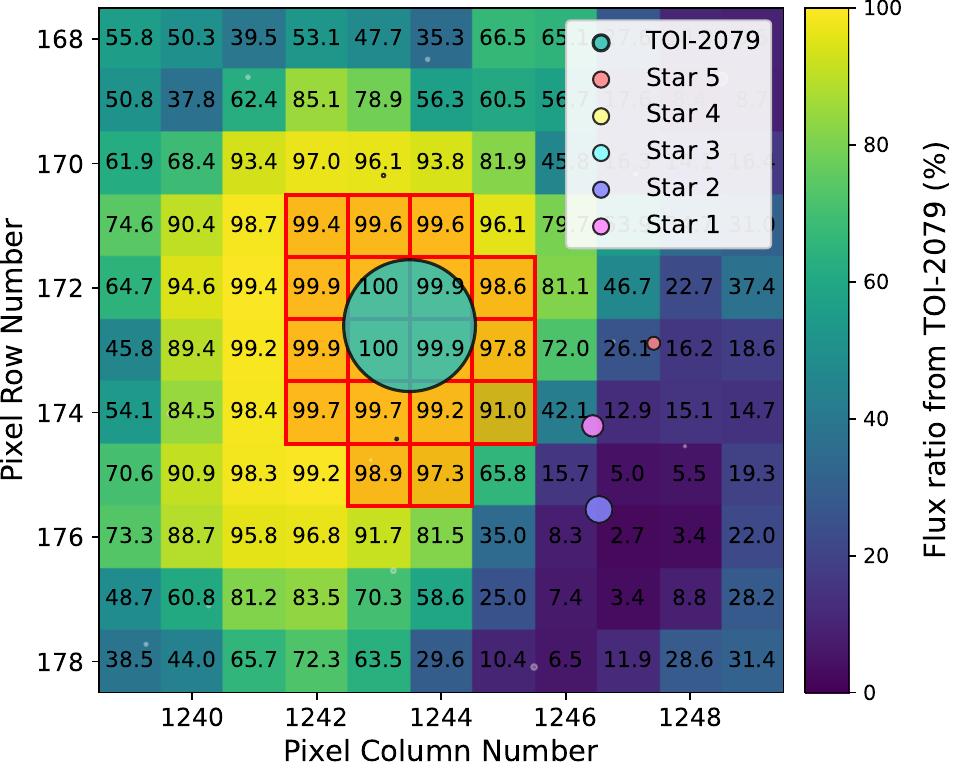}
    \includegraphics[width=0.19\linewidth]{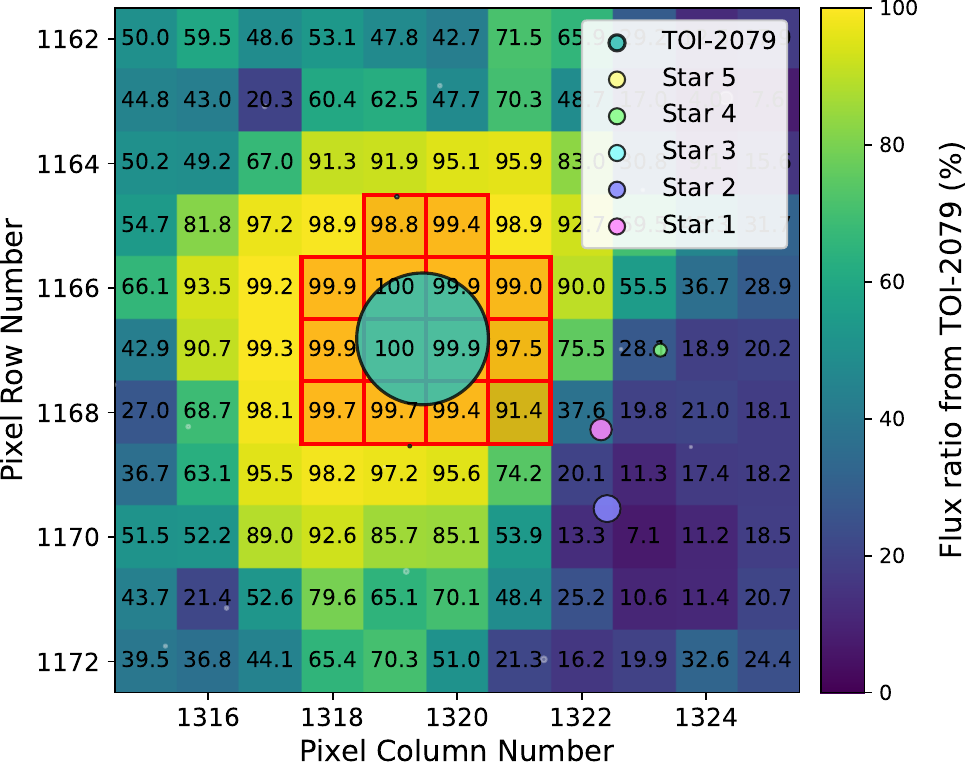}
    \caption{All TESS sectors' contaminating sources within the SPOC aperture (red mask) overlaid on the heat map from TESS observations of the target star, TOI-2079. The sizes of the target and nearby stars are scaled according to their relative fluxes in the TESS filter. The pixel scale is $21\arcsec$ per pixel. This plot was generated using \texttt{TESS-cont} \citep{2024A&A...691A.233C}.}
    \label{fig:tpfs}
\end{figure*}

\begin{figure}
    \centering
    \includegraphics[width=\linewidth]{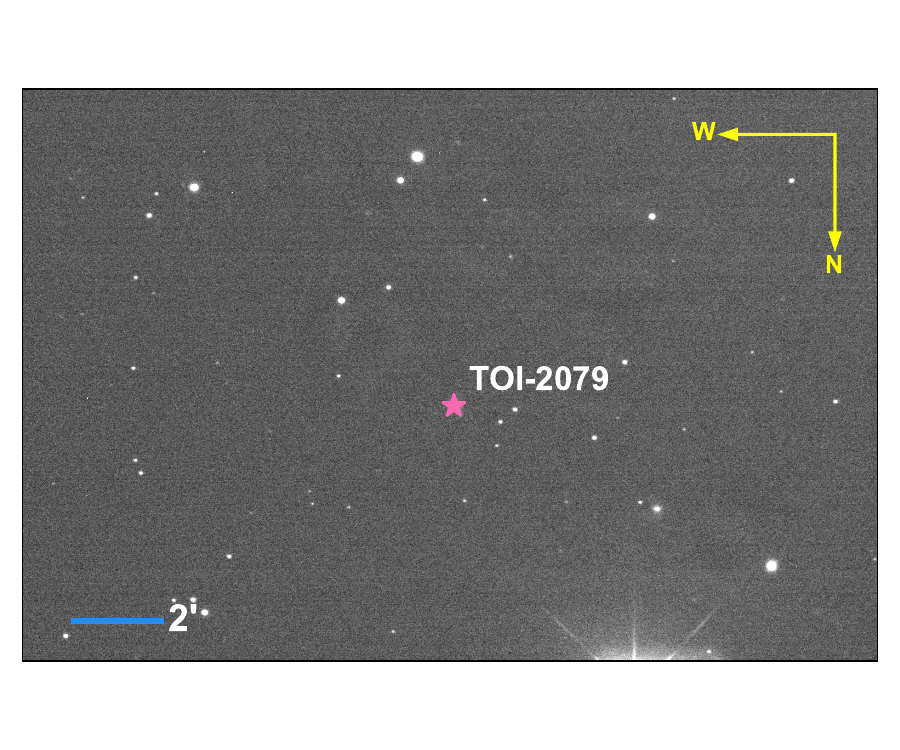}
    \caption{TOI-2079's field image taken from the OAF on April $26$, $2026$. TOI-2079 is a northern target, and its field does not appear to be particularly crowded, with no obvious indication of severe source confusion.}
    \label{fig:field_oaf}
\end{figure}

In this Appendix we show the TESS target pixel files (Figure \ref{fig:tpfs}), the Observatorio Astronómico de Forcarei (OAF; Galicia, Spain) field image (Figure \ref{fig:field_oaf}), and we also describe the ground-based photometric observations used for the validation of TOI-2079\,b and TOI-2079\,c, which are summarized in Table~\ref{tab:observations}. Following the approach of \cite{2023A&A...672A..70P}, we have considered for analysis only those observations with a Bayes factor $\Delta\ln Z\,>\,5$ in order to preserve the quality of the study (see Table\,\ref{tab:observations}), where $\Delta\ln Z\,=\,\ln Z_{\mathrm{transit}}\,-\,\ln Z_{\mathrm{noise}}$. Therefore, the ground-based observations used to validate TOI-2079.01 and TOI-2079.02 are the MuSCAT2 observation obtained on March $3$, $2021$, and the \textit{Las Cumbres Observatory} (LCO) observation obtained on December $12$, $2025$, respectively.

\begin{table*}
    \centering
    \caption{TOI-2079 ground-base observations.}
    \begin{tabular*}{\textwidth}{@{\extracolsep{\fill}}c c c c c c c}
    
        \hline\hline
        \noalign{\smallskip}
        TIC & Observatory & Telescope & Instrument & Filter & Date & Bayes Factor \\
        
        \noalign{\bigskip}
        TOI-2079.01 &  &  &  &  &  &  \\
        \hline
        
        \noalign{\smallskip}
        224298134.01 & \textit{Las Cumbres} (Teid) & $1$\,m & Sinistro & $ip$ & $2022$\,Dic\,$16$ & $<\,5$ \\
         & \textit{Las Cumbres} (McD) & $1$\,m & Sinistro & $ip$ & $2021$\,Apr\,$25$ & $<\,5$ \\
         &  & $1$\,m & Sinistro & \textit{ip} & $2021$\,Apr\,$13$ & $<\,5$ \\
         &  & $1$\,m & Sinistro & \textit{ip} & $2020$\,Dic\,$16$ & $<\,5$ \\
         &  & $1$\,m & Sinistro & \textit{ip} & $2020$\,Dic\,$16$ & $<\,5$ \\
         & \textit{Teide} & \textit{Carlos Sánchez} & MuSCAT2 & \textit{g}, \textit{r}, \textit{i}, \textit{z} & $2021$\,Mar\,$03$ & $6.94\dagger$, $5.73\dagger$, $8.99\dagger$, $7.84\dagger$ \\
         
        \noalign{\bigskip}
        TOI-2079.02 &  &  &  &  & \\
        \hline
        
        \noalign{\smallskip}
        224298134.02 & \textit{Las Cumbres} (McD) & $1$\,m & Sinistro & \textit{ip} & $2025$\,Dec\,$12$ & $5.38\dagger$ \\
         & \textit{Canis Major} & $0.4$\,m & SBIG STXL & \textit{IC} & $2025$\,Apr\,$04$ & $<\,5$ \\
         & \textit{TÜBİTAK} & T100\,($1$\,m) & SI $1100$ & \textit{zp} & $2024$\,Feb\,$19$ & $<\,5$ \\
         & \textit{San Pedro Mártir} & SAINT-EX\,($1$\,m) & SI $1100$ & \textit{zp} & $2024$\,Feb\,$19$ & $<\,5$ \\
        \hline
        
    \end{tabular*}
    \tablecomments{All Observations with Bayes factor values $\Delta$\,ln\,Z greater than $5$ are highlighted with $\dag$ (see Section\,\ref{sec:observations_p_appendix}).}
    \label{tab:observations}
\end{table*}

\subsection{\upshape{Tierras} \textit{observations}}\label{subsec:Tierras}
We observed TOI-2079 with the 1.3-m \textit{Tierras} Observatory \citep{GarciaMejia2020} on 25 nights over an almost 60-day baseline (see Figure~\ref{fig:tierras}). Observations were taken at a 60-s exposure time in the custom \textit{Tierras} filter, a narrow-band filter in the near-infrared that was designed to avoid telluric contamination ($\lambda_C = 863.5$~nm; FWHM~=~40.2~nm). The images were processed and light curves produced with the \textit{Tierras} photometric pipeline \citep{Tamburo2025}. 

Analysis of the \textit{Tierras} light curve with a Lomb-Scargle (LS) periodogram revealed a strong periodicity at 17.3~days and an amplitude of about 1\%. We examined fits of sinusoidal models with periods corresponding to the less significant peaks in the periodogram to rule out aliasing of the 17.3-day signal, and found that the other periodicities yield significantly worse fits. To estimate the uncertainty on the 17.3-day signal, we then fit quasi-periodic Gaussian process (QP-GP) models to the \textit{Tierras} light curve. The covariance function of these models, as implemented in \texttt{celerite} \citep{ForemanMackey2017}, take the form 

\begin{equation}
    \kappa(\tau) = \frac{B}{2+C} e^{-\tau/L} \left[ \cos\left( \frac{2 \pi \tau}{P_{\mathrm{rot}}}\right) + (1+C)\right] + \sigma^2\delta_{\mathrm{ij}}.
\end{equation}

Here, $\tau$ is a matrix of the absolute differences between each pair of time stamps of the light curve in days. $P_\mathrm{rot}$ represents the rotation period of target, and $L$ is an exponential decay timescale, both in days. $B$ controls the amplitude of the covariance (in units of ppt$^2$), while $C$, which is unitless, controls the relative strength of the periodic and aperiodic terms of the kernel. We have additionally included a jitter term, $\sigma^2\delta_\mathrm{ij}$, to account for potentially underestimated error bars.  

We sampled the log of the hyperparameters of the QP-GP kernel using $\texttt{emcee}$ \citep{ForemanMackey2013}. We present the bounds that we applied to the hyperparameters in Table~\ref{tab:gp_hyperparams}. The upper bound on $B$ corresponds to the measured variance of the \textit{Tierras} light curve. The lower bound on $L$ permits decay timescales as short as four times the time span of the light curve, since the variability amplitude does not evolve significantly over the duration of the observations. We ran 100 walkers until the chains were longer than 50$\times$ the autocorrelation time, which required 6,700 steps, and we discarded the first 10\% of these chains for burn-in. In Figure~\ref{fig:tierras}, we visualize the model with the maximum log probability, along with its 1- and 2-$\sigma$ uncertainty contours. The posterior distribution of $P_\mathrm{rot}$ has a median value and 16\textendash 84 percentile range of \Prot~days, which we adopt as the rotation period of TOI-2079.

\begin{figure*}
    \centering
    \includegraphics[width=\textwidth]{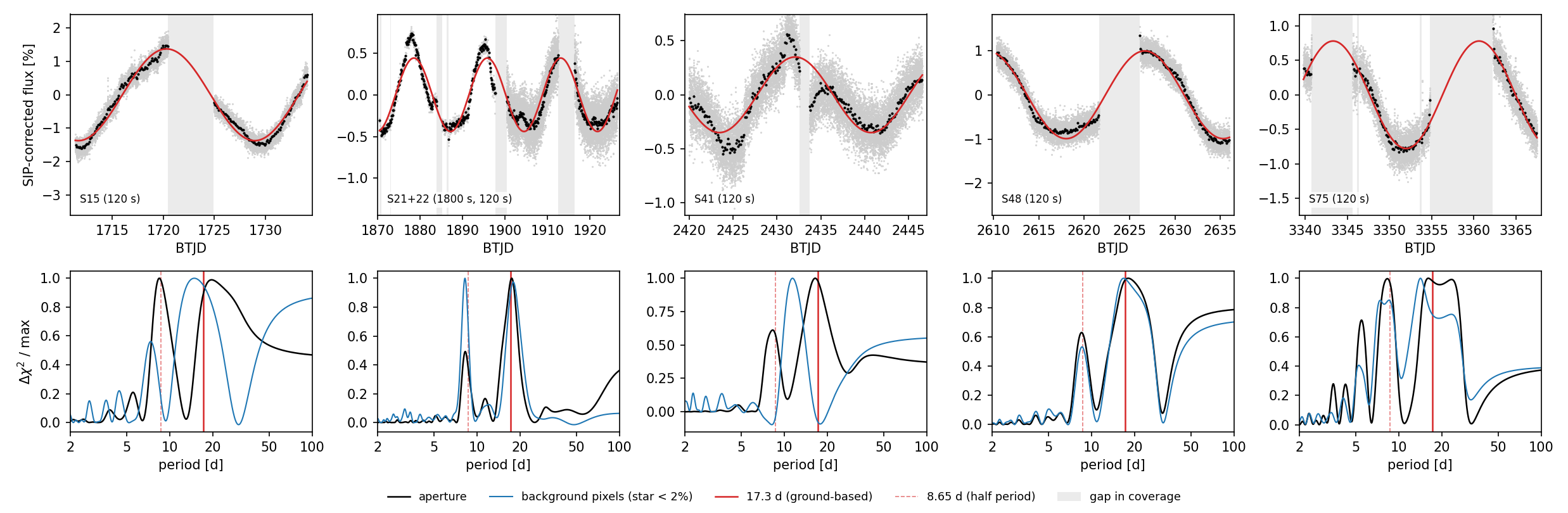}
    \caption{TESS photometry (top row) and periodograms (bottom row) for TOI-2079 in Sectors 15, 21+22, 41, 48, and 75. We re-reduced these data with the systematics-insensitive periodogram method \citep{Hedges2020}, finding strong evidence for an approximately 17.5-day signal independently in each Sector. This is consistent with the \textit{Tierras} measurement of \Prot~days. However, periodograms of the flux in the background pixels (blue in bottom row) also show significant power at this period in most cases, highlighting the importance of the ground-based data for uniquely associating the period with TOI-2079.}
    \label{fig:TESS_periodograms}
\end{figure*}

\begin{figure}
    \centering
    \includegraphics[width=\columnwidth]{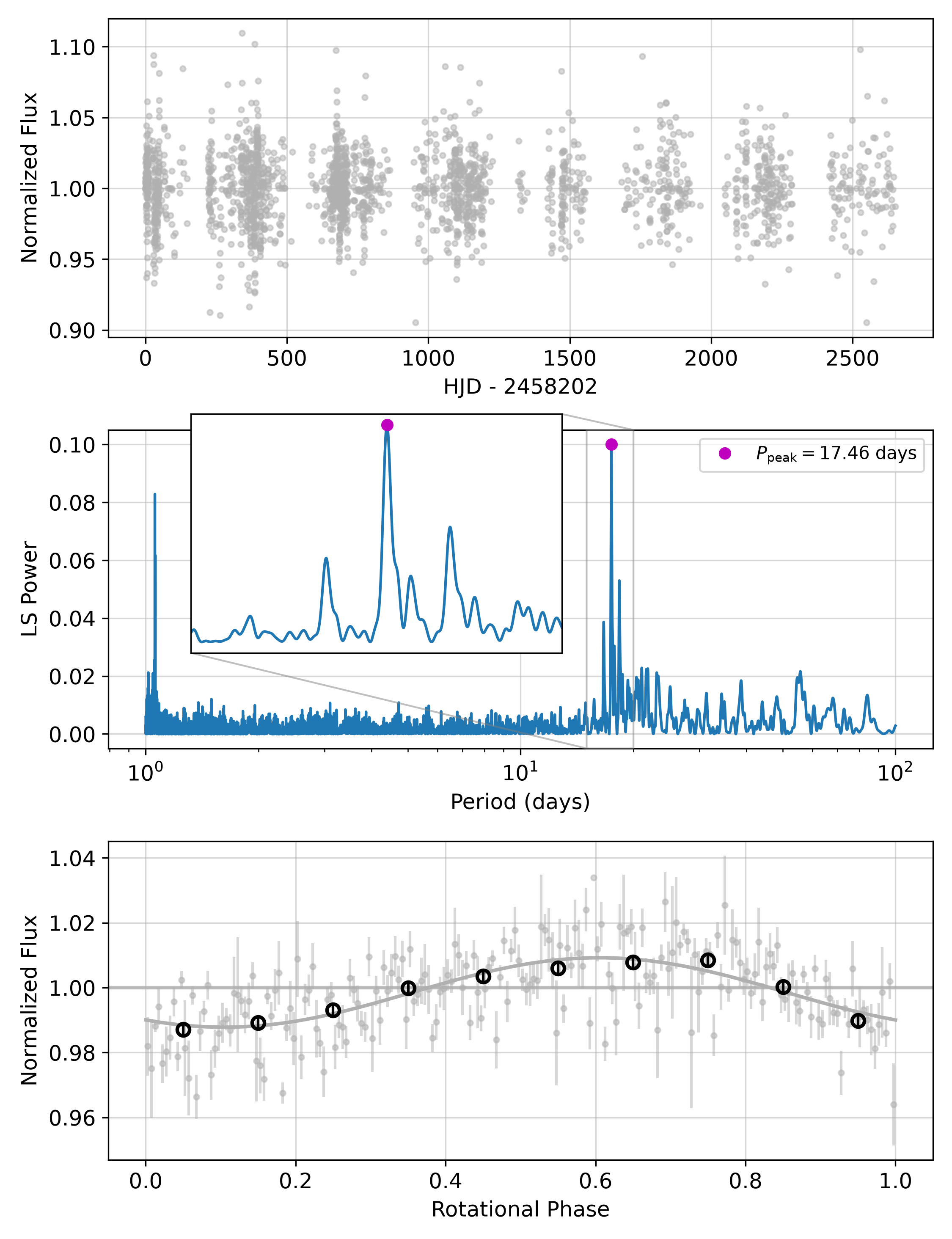}
    \caption{Top row: $g$-band photometry from ZTF for TOI-2079, spanning March 2018 to June 2025. These data were taken with Cameras 13 and 16, and we median filtered both with a window of 75 days to remove camera and seasonal offsets before combining them into one light curve. Middle row: The LS periodogram of the data, revealing a strong peak at 17.46~days. Bottom row: The ZTF data phase-folded on a period of 17.46~days. The gray points in this panel show the phased data in phase bins of width 0.005, while the black points show the data in phase bins of width 0.1, to more clearly show the variability signal.}
    \label{fig:ztf}
\end{figure}

To verify that the rotation period is stable over long timescales, we re-reduced the TESS observations and searched for variability using the systematics-insensitive periodogram method \citep{Hedges2020}. Additionally, we ran a Lomb-Scargle periodogram on $g$-band photometry of TOI-2079 from the Zwicky Transient Factory \citep[ZTF;][]{Bellm2019} spanning from March 2018 to June 2025 after performing a 75-day median filtering to correct for season-to-season offsets. We found evidence for strong, stable 17.5~d periodicity in both datasets (see Figures~\ref{fig:TESS_periodograms} and \ref{fig:ztf}), indicating that TOI-2079 has maintained the rotation period observed by \textit{Tierras} for nearly a decade. Since the \textit{Tierras} light curve has much higher per-exposure SNR than the ZTF data and has more than double the time baseline of a single 27-day TESS sector, we elect to interpret the 17.33-day signal from the \textit{Tierras} light curve as the true rotation period of the star.

\begin{deluxetable}{ll} \label{tab:gp_hyperparams}
\tablecaption{Bounds on the QP-GP hyperparameters.}
\tablehead{\colhead{Parameter} & \colhead{Prior} }
\startdata
$\ln(B/\mathrm{ppt}^2)$ & $\mathcal{U}(-20.0,3.2)$ \\
$\ln(C)$ & $\mathcal{U}(-20.0,20.0)$ \\
$\ln(L/\mathrm{day})$ & $\mathcal{U}(5.4, 20.0)$\\
$\ln(P_\mathrm{rot}/\mathrm{day})$ & $\mathcal{U}(-20.0, 20.0)$\\
$\ln(\sigma^2/\mathrm{ppt}^2)$ & $\mathcal{U}(-20.0, 20.0)$\\
\enddata
\end{deluxetable}

\subsection{MuSCAT2}
\label{subsec:observations_gbp_appendix_muscat2}

We observed one transit of TOI-2079.01 with the Multicolour Simultaneous Camera for studying Atmospheres of Transiting exoplanets \citep[MuSCAT2;][]{2019JATIS...5a5001N} mounted on the Telescopio \textit{Carlos Sánchez} (TCS) at the Observatorio del Teide (OT), Spain. This instrument is a multi-colour imager able to perform simultaneous photometry in four photometric bands (Sloan-$g$, -$r$, -$i$, and $z_\mathrm{s}$) using four separate CCDs. Each channel has independent exposure times. Usually, when observing M dwarfs, the g filter's CCD is set to have shorter than $15$\,s exposure times and this channel’s images are normally used to auto-guide the instrument.

The MuSCAT2 data were reduced using a dedicated pipeline \citep{2020A&A...633A..28P}, which optimized comparison star selection and aperture sizes for extracting the target’s photometric time series.

\subsection{LCOGT-1m0}
\label{subsec:observations_gbp_appendix_lco}

We observed five transits of TOI-2079.01 and one of TOI-2079.02 with the Las Cumbres Observatory Global Telescope \citep[LCOGT;][]{2013PASP..125.1031B} $1$\,m network nodes installed in the \textit{McDonald} Observatory and OT. The $1$\,m telescopes are equipped with $4096\times4096$ SINISTRO cameras, with an image scale of $0.389\arcsec$ per pixel, resulting in a $26\arcmin\times26\arcmin$ field of view. The observations were conducted simultaneously in the Sloan-$g'$, -$r'$, -$i'$, and Pan-STARRS-$z$ filters. Science images calibration and photometric extraction were performed using the LCOGT {\tt BANZAI} pipeline \citep{2018SPIE10707E..0KM} and {\tt AstroImageJ}, respectively.

\subsection{Canis Major $0.4$\,m}
\label{subsec:observations_gbp_appendix_canismajor}


We observed a full transit of TOI-2079.02 on April $4$, $2025$ (UT) in the $ic$ filter with the 0.4\,m telescope of the Canis Major Observatory (CMO) in Liguria, Italy. The telescope is equipped with a 3072 $\times$ 2048 SBIG STXL-6303E camera with an image scale of
1$\arcsec$.16 pixel$\rm^{-1}$ resulting in a $49\arcmin \times 33\arcmin$ field of view. The images
were calibrated and the photometric data were extracted using {\tt AstroImageJ} \citep[AIJ; ][]{Collins2017}.

\subsection{TÜBİTAK T100}
\label{subsec:observations_gbp_appendix_tubitak}

We observed a partial transit of TOI-2079.02 on February $19$, $2024$ (UT) in the $z'$ filter with the 1\,m Turkish telescope (TUG100). The telescope is located at the Antalya campus of the Türkiye National Observatories (TUG) at an altitude of $2500$ meters. It is equipped with a cryo-cooled SI $1100$ CCD featuring $4096\times4096$ pixels and a field of view of $21\arcmin \times21\arcmin$, and was operated in $2\times2$ binning mode. The data reduction and differential photometry performed relative to the summed flux of comparison stars were carried out using {\tt AstroImageJ} \citep[AIJ; ][]{Collins2017}.

\section{Spectroscopic observations}
\label{sec:observations_s_appendix}

\subsection{IRTF/SpeX}
\label{subsec:spectroscopy_appendix_spex}

Observations were obtained in the short-wavelength cross-dispersed (SXD) mode using the $0\farcs3 \times 15''$ slit aligned to the parallactic angle, providing $R{\sim}2000$ spectra over 0.80--2.42\,$\mu$m.
Six 90-s exposures at an airmass of 1.5 were collected while nodding in an ABBA pattern.
Standard SXD flat-field and arc-lamp calibration frames were acquired, along with six 30-s exposures of the A0\,V standard HD\,99966 ($V{=}7.1$) at a similar airmass.
We reduced the data with the pySpextool pipeline\footnote{\url{https://github.com/pyspextool}}, based on the Spextool v4.1 pipeline \citep{Cushing2004}.
The final spectrum has a median SNR of ${\sim}200$ per resolution element.

\begin{figure}
    \centering
    \includegraphics[width=\linewidth]{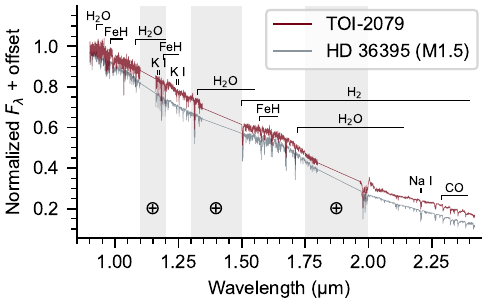}
    \caption{
    SpeX SXD spectrum of TOI-2079.
    The target spectrum (red) is shown along with that of the M1.5V standard HD\,36395 (grey), offset for clarity.
    Prominent atomic and molecular absorption features in M-dwarf spectra are labeled, and regions strongly affected by telluric absorption are shaded.
    }
    \label{fig:spex}
\end{figure}

\subsection{Shane/Kast}
\label{subsec:spectroscopy_appendix_kast}

We used the 1$\arcsec$ slit to obtain blue and red optical spectra split at 5700~{\AA}, and the 600/4310 grism in blue and 600/7500 grating in red, resulting in spectral resolutions of $\lambda/\Delta\lambda$ $\approx740$ at 5425~{\AA} and $\lambda/\Delta\lambda$ $\approx980$ at 7450~{\AA}. 
We obtained one 200\,s exposure in blue and two 90\,s exposures in red at an average airmass of 1.14. 
The spectrophotometric calibrator Hiltner 600 \citep{1992PASP..104..533H,1994PASP..106..566H} was observed at the start of the night for flux calibration, and the G2~V star HD~117845 ($V$=8.1) was observed immediately after TOI-2079 for telluric absorption calibration.  
HeNeArHg arc lamp, flat-field lamp, and bias frame exposures were obtained at the start of the night for wavelength calibration, pixel-response correction, and bad-pixel masking. 
Data were reduced using the kastredux code\footnote{\url{https://github.com/aburgasser/kastredux}.} version 1.1.4 \citep{kastredux} using standard settings. 
Due to observations being obtained close to sunrise, the spectral signal-to-noise (S/N) was relatively low at 21 at 5425~{\AA} and 26 at 7450~{\AA}, and portions of the red spectrum were masked due to poor telluric correction and strong scattered light contributions.

\begin{figure}
    \centering
    \includegraphics[width=\linewidth]{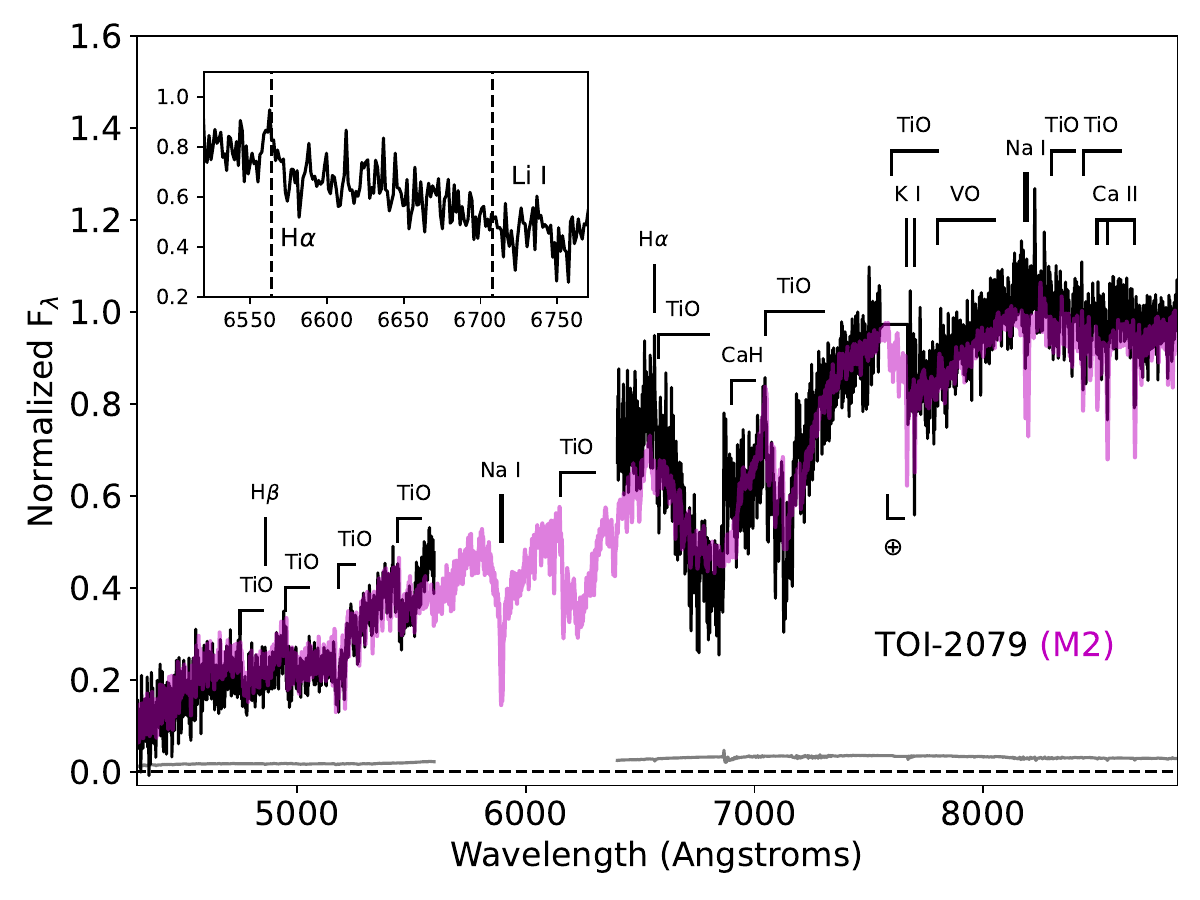}
\caption{Normalized Shane/Kast optical spectrum of TOI-2079, with flux density shown as the black line and uncertainty as the grey line. This is compared to the best-fit M2 SDSS spectral template from \citet[magenta line]{2007AJ....133..531B}. Key spectral features are indicated, including a masked region of strong telluric absorption ($\oplus$).
The gap in data between 5600~$\AA$ and 6400~$\AA$ corresponds to the gap between the blue and red channels of Kast, as well as the red spectrum affected by strong scattered light. 
The inset box highlights the 6520--6770~{\AA} region, which includes a weak 6563~{\AA} H$\alpha$ emission feature and absent 6708~{\AA} Li\,\textsc{i} absorption.}
    \label{fig:kast}
\end{figure}

\subsection{HST/STIS}
\label{subsec:spectroscopy_appendix_hst}

We obtained the far-ultraviolet spectrum of TOI-2079 with the Hubble Space Telescope's (HST) Space Telescope Imaging Spectrograph (STIS) as part of the Survey of Transiting Exoplanets in Lyman-alpha (STEL$\alpha$; HST-GO-17804; PIs: Parke Loyd \& Vissapragada). One science exposure ($\sim3000$s) was taken on 2025 March 23 using the $0.2$\arcsec~x $0.2$\arcsec~aperture. We used the far-ultraviolet multi-anode microchannel array configuration (FUV-MAMA) and the E140M echelle grating (R$\sim 45800$) covering $1144-1729$ \AA. We manually located the spectral trace and extracted the target's spectrum using the {\tt stistools} pipeline.

Due to absorption from the interstellar medium (ISM) and contamination from the Earth's geocorona, the core of the Lyman-$\alpha$ line is obstructed. We used the {\tt lyapy} package to reconstruct TOI-2079's intrinsic Lyman-$\alpha$ emission \citep{2016ApJ...824..101Y,2022zndo...6949067Y}. We modeled the emission line as a Voigt profile described by the line center, integrated line flux, and Gaussian and Lorentzian widths. We included a self-reversal parameter. The ISM absorption was modeled as a single-cloud component Voigt profile parameterized by neutral H column density, Doppler broadening, neutral H velocity, and the deuterium-to-hydrogen ratio (D/H). We fixed D/H to $1.5\times 10^{-5}$, consistent with the literature \citep{2006ApJ...647.1106L}. Details on the Lyman-$\alpha$ data reduction, extraction, reconstruction, and results for all STELa targets will be presented in Rockcliffe, et. al. (in prep).

\section{High-resolution Imaging}
\label{sec:hr_imaging_appendix}

As part of our standard process for validating transiting exoplanets to assess the possible contamination of bound or unbound companions on the derived planetary radii \citep{ciardi2015}, we observed TOI-2079 with near-infrared adaptive optics imaging on Palomar and optical speckle imaging at SAI. The near-infrared and optical imaging complement each other, offering different resolutions and sensitivities.

\subsection{Palomar NIR Adaptive Optics}
	Observations of TOI-2079 were made on 2021-Jun-21 with the PHARO instrument \citep{hayward2001} on the Palomar Hale (5m) behind the P3K natural guide star AO system \citep{dekany2013}. The pixel scale for PHARO is $0.025\arcsec$. The Palomar data were collected in a 3--point quincunx dither pattern in the $K_{cont}$ filter. The reduced science frames were combined into a single mosaiced image with final resolutions of $\sim 0.10$\arcsec and $\sim 0.12$\arcsec, respectively.
	
	The sensitivity of the final combined AO images was determined by injecting simulated sources azimuthally around the primary target every $20^\circ $ at separations of integer multiples of the central source's FWHM \citep{furlan2017}. The brightness of each injected source was scaled until standard aperture photometry detected it with $5\sigma $ significance.  The final $5\sigma $ limit at each separation was determined from the average of all of the determined limits at that separation, and the uncertainty on the limit was set by the rms dispersion of the azimuthal slices at a given radial distance.

\subsection{Shane NIR Adaptive Optics}
\label{subsec:hr_imaging_shane_appendix}

TIC 224298134 (TOI-2079) was observed on UT 2020 December 1 using the ShARCS camera on the Shane 3\,m telescope at Lick Observatory \citep{2012SPIE.8447E..3GK, 2014SPIE.9148E..05G, 2014SPIE.9148E..3AM}. Observations were taken with the Shane adaptive optics system in natural guide star mode in order to search for nearby, unresolved stellar companions. They collected a single sequence of observations using a $Ks$ filter ($\lambda_0 = 2.150$ $\mu$m, $\Delta \lambda = 0.320$ $\mu$m). The data were reduced using the publicly available \texttt{SImMER} pipeline \citep{2020AJ....160..287S, 2022PASP..134l4501S}.\footnote{https://github.com/arjunsavel/SImMER} No nearby stellar companions within detection limits. More information about these observations can be found in Dressing et al. (submitted).

\subsection{Gemini Optical Speckle}
\label{subsec:hr_imaging_gemini_appendix}

TOI-2079 was observed on UT 2021 February 3 with the 'Alopeke speckle instrument on the 8.1\,m Gemini North telescope on Mauna Kea, Hawaii \citep{2021FrASS...8..138S}. The target was observed simultaneously in two narrow-band filters, centered at 562 and 832 nm. Eight sets of 1000 x 0.06 s exposures were obtained and reduced using the standard reduction people (see \cite{2011AJ....142...19H}). The output data products include reconstructed images and 5-sigma contrast curves to provide limits on companion detections. We do not detect any nearby companions with a differential magnitude between ~4.5 and 7.5 mag from 0.1" to 1.2", and we can exclude brighter companions down to the diffraction limit of the telescope (20 mas).

\subsection{SAI Optical Speckle}
\label{subsec:hr_imaging_sai_appendix}

We observed TOI-2079 on December 2nd, 2020, with the speckle polarimeter on the 2.5\,m telescope at the Caucasian Observatory of Sternberg Astronomical Institute (SAI) of Lomonosov Moscow State University \citep{Safonov2017}. An electron-multiplying CCD Andor iXon 897 was used as a detector. The atmospheric dispersion compensator was active, which allowed the use of the $I_\mathrm{c}$ band. The respective angular resolution is 0.083''. 5940 frames with an exposure of 29.7 ms were accumulated. No stellar companions were detected, $5\sigma$ limits at distances $0.25$ and $1.0^{\prime\prime}$ from the star are $\Delta=4.1^m$ and $6.9^m$.

\begin{figure}
    \centering
    \includegraphics[width=\linewidth]{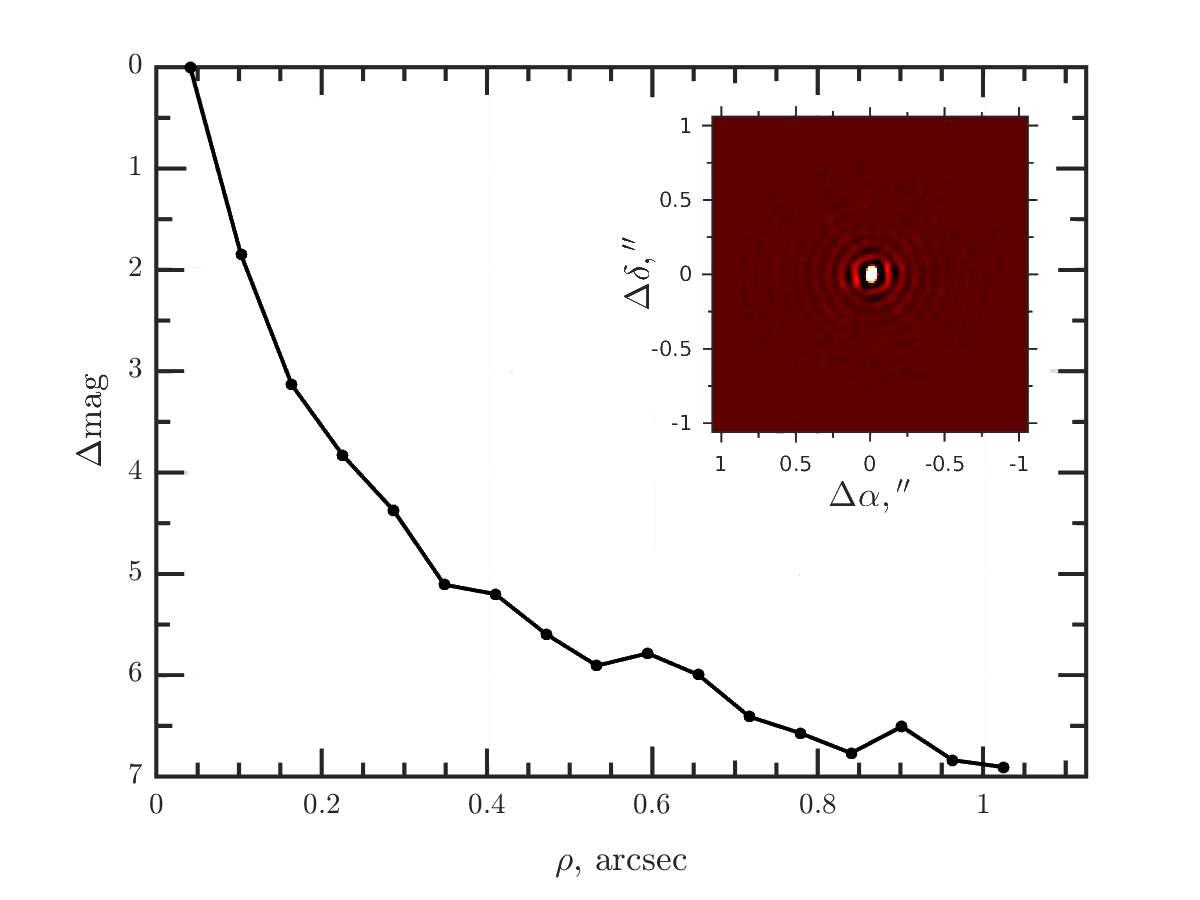}
    \caption{High-resolution imaging and contrast curves as a function of angular separation of TOI-2079 obtained with the Speckle Polarimeter installed at the $2.5$\,m telescope at Caucasian Observatory of SAI.}
    \label{fig:sai-2.5m}
\end{figure}

\section{Transit analysis}
\label{sec:transit_analysis_appendix}

We used the \texttt{SHERLOCK} pipeline \citep{pozuelos2020,devorapajares2024} to explore the available \textit{TESS} photometry of TOI-2079 and search for additional transiting planet candidates that may have gone unnoticed by the official SPOC and QLP pipelines. \texttt{SHERLOCK} is an end-to-end, open-source software designed to recover known TESS Objects of Interest (TOIs) and to identify new planetary candidates through a systematic, iterative analysis of space-based light curves. The pipeline comprises six modules that sequentially handle the acquisition and pre-processing of photometric data, the detection of transit-like signals via the Transit Least Squares (TLS) algorithm \citep{tls}, the vetting of each detected signal using \texttt{WATSON} \citep{devorapajares2024} to reject instrumental artifacts and astrophysical false positives, the statistical validation of candidates, the joint Bayesian modeling of the system architecture with \texttt{ALLESFITTER} \citep{allesfitter}, and the computation of observational windows for ground-based follow-up. For each bandpass, we computed the LD coefficients using \texttt{ExoTETHyS} \citep{Morello2020joss,Morello2020} in combination with \texttt{PHOENIX-COND} stellar atmosphere models \citep{Husser2013,Claret2018}.

For TOI-2079, we downloaded all available \textit{TESS} 2-minute cadence Pre-search Data Conditioning Simple Aperture Photometry (PDCSAP) light curves via \texttt{LIGHTKURVE} \citep{lightkurve}. \texttt{SHERLOCK} then performed a blind transit search over the full light curve using the available short-cadence sectors ($15$, $22$, $41$, $48$, $75$) with an iterative approach: the strongest periodic signal was identified with TLS, its corresponding transits were masked, and the search was repeated on the residuals until no additional signal exceeded the signal-to-noise detection threshold. Each detected signal was cross-matched against the official TOI catalogue to distinguish previously alerted candidates from potential new discoveries, and was subsequently passed through the \texttt{WATSON} vetting module to assess its planetary nature against common false-positive scenarios.

In the case of TOI-2079, \texttt{SHERLOCK} successfully recovered both known transiting planets---TOI-2079\,b ($P\sim1.49$\,d) and TOI-2079\,c ($P\sim9.32$\,d)---at high signal-to-noise ratio. No additional statistically significant transit signals were detected in the residual light curve after masking these two planets, indicating that the available \textit{TESS} photometry shows no evidence for further transiting companions in this system down to the detection sensitivity of the pipeline.

In addition to the periodic transit search, we employed \texttt{MORIARTY}, a single-transit detection module within the \texttt{SHERLOCK} framework, to search for isolated transit events that could be produced by long-period planets with only one transit captured during the \textit{TESS} observing window. \texttt{MORIARTY} uses a convolutional neural network trained on synthetic transit injections, to compute a transit-likelihood score for every cadence in the light curve. Regions exceeding a probability threshold of $0.5$ are flagged as candidate single-transit events and subsequently fitted with a box-shaped transit model to estimate their depth, duration, and epoch. For each candidate, an orbital period distribution is inferred through a Monte Carlo sampling of the single-transit geometric degeneracy, and a projected signal-to-noise ratio check is applied to reject events whose estimated period would have produced multiple detectable transits in the available data. Candidates are also filtered against the Roche limit estimating their period likelihood based on their transit duration, and inspected for contamination by momentum dumps, data gaps and background flux differences. We ran \texttt{MORIARTY} on the full \textit{TESS} light curve of TOI-2079 after masking the transits of planets b and c. While a small number of low-significance single-transit-like features were identified by the neural network, none passed the full set of vetting criteria: the surviving candidates either had nonphysically short estimated orbital periods below the Roche limit, or their inferred periods would have rendered them detectable through the standard phase-folded transit search, indicating they are most likely spurious detections or residual systematic artifacts. We therefore conclude that no reliable single-transit events are present in the current \textit{TESS} data for this system.

\section{Validation of TOI-2079 system}
\label{sec:validation_appendix}

The validation of transiting exoplanets involves assessing the probability of astrophysical false-positive scenarios mimicking planetary signals. Several statistical tools have been developed for this purpose, including \texttt{BLENDER} \citep{2011ApJ...727...24T}, \texttt{VESPA} \citep{2012ApJ...761....6M, 2016ApJ...822...86M}, and \texttt{TRICERATOPS} \citep{2021AJ....161...24G}. These methods combine transit photometry and stellar constraints to estimate the false-positive probability (FPP), with values below 1\% commonly adopted as the threshold for statistical validation.

Throughout the wavelength range covered by the filters used for TOI-2079.01 and TOI-2079.02, we obtained the following values of $R_{\rm p}/R_{\star}$: $0.0248^{+0.0013}_{-0.0018}$, $0.02606^{+0.0010}_{-0.0011}$, $0.0234^{+0.0016}_{-0.0018}$, $0.02564^{+0.0010}_{-0.0013}$, and $0.0287^{+0.0019}_{-0.0029}$ for the MuSCAT2 \textit{g}, \textit{i}, \textit{r}, \textit{z}, and LCO \textit{ip} bands, respectively. As no significant chromaticity is detected for either planet, we proceed with the subsequent validation steps, as described below.

\begin{figure}
    \centering
    \includegraphics[width=\linewidth]{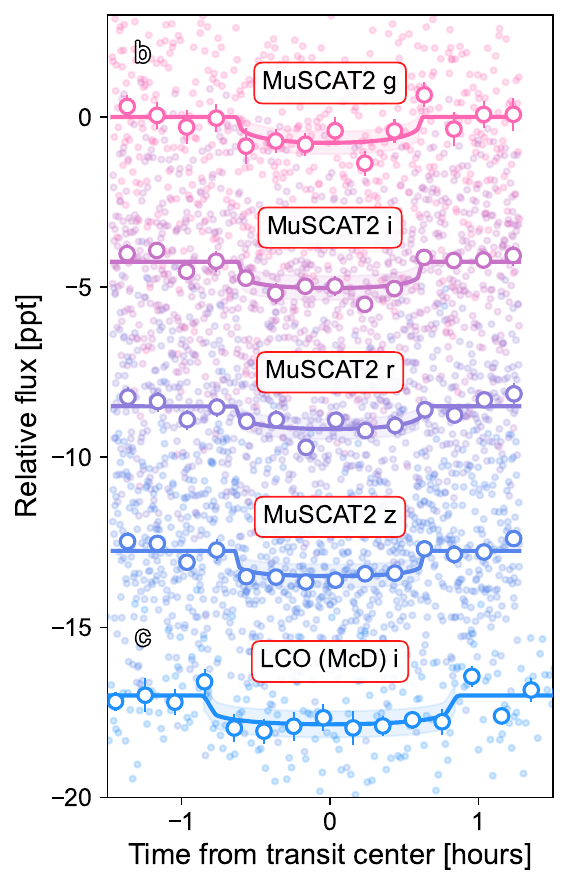}
    \caption{MuSCAT2 and LCO (McD) light curves for TOI-2079.01 and TOI-2079.02 respectively, together with the transit models (colored line) showing the 1$\sigma$ range (colored region) in models during transit. All observed data is binned in phase in steps of $12$ minutes.}
    \label{fig:gb_photometry}
\end{figure}

\subsection{Statistical validation}
\label{subsec:statistical_validation_appendix}

This approach estimates the relative likelihoods of different astrophysical configurations capable of producing the observed transit-like photometric signal, by combining \textit{TESS} light curves, Gaia astrometric and photometric information, contrast curves from high-resolution imaging, and Galactic stellar population models. The \texttt{TRICERATOPS} framework distinguishes between validated planets, likely candidates, and nearby false positives, relying on two parameters: the FPP and the Nearby False Positive Probability (NFPP). The FPP quantifies the likelihood that the observed event is not of planetary origin, while the NFPP measures the probability that the transit-like signal originates from a nearby contaminant source. Following the standard classification criteria, systems with FPP\,$<$\,0.015 and NFPP\,$<$\,0.001 are considered as validated planets, those with FPP\,$<$\,0.5 and NFPP\,$<$\,0.001 are likely planetary candidates, and cases with NFPP\,$>$\,0.1 are treated as probable nearby false positives. For our analysis, we employed the \textit{TESS} photometry described in Section~\ref{sec:observations}, together with high-resolution imaging contrast curves from Section~\ref{sec:hr_imaging_appendix}.

\subsection{Neural network-based statistical vetting with \texttt{WATSON-Net}}
\label{subsec:validation_ml}
As an independent layer of vetting complementing the \texttt{TRICERATOPS} statistical validation, we submitted both planet candidates to the \texttt{WATSON} vetting engine \citep{devorapajares2025}, which ingests the \texttt{SHERLOCK} \citep{devorapajares2024} prepared data and a set of \texttt{ALLESFITTER}-derived \citep{allesfitter} transit parameters to feed them into a deep neural network (\texttt{WATSON-Net}) trained to discriminate between transiting planets and astrophysical and instrumental false positives. The network evaluates several individual model inputs including the odd-even transit view, secondary eclipse view, centroid shifts view, optical ghost diagnosis view, goodness-of-transit statistics, simplistic bootstrap false-alarm probability (FAP), and photometric stellar and transit parameters, and aggregates them into a calibrated planetary probability score.

For planet b ($P\,=\,1.49$\,d), \texttt{WATSON-Net} places it firmly in the high-confidence planetary regime. All diagnostic tests passed cleanly for this candidate: the secondary eclipse signal is negligible, the odd-even depth difference is consistent with zero at $0.32\sigma$, the centroid offset signal-to-noise ratios show no evidence of source contamination, and the simplistic bootstrap FAP is zero. The \texttt{WATSON-Net} prediction thus independently reinforces the statistical validation and strongly supports the planetary nature of the inner signal.

For planet c ($P\,=\,9.32$d), \texttt{WATSON-Net} yielded a slightly lower calibrated planetary probability with a larger uncertainty, primarily due to the smaller number of observed transits for this longer-period candidate, which reduces the constraining power of the odd--even and centroid diagnostics. Nevertheless, the core metrics remain reassuring: the secondary eclipse significance is low ($SNR\,=\,0.31$), the odd–even depth difference is consistent with zero at $0.64\sigma$, the bootstrap FAP is zero, and the optical ghost diagnosis and source-offset checks are consistent with an on-target transit source. While the \texttt{WATSON-Net} confidence does not reach the same level as for planet b, no individual branch of the network decisively flags a false-positive scenario, and the combined evidence from \texttt{WATSON-Net} and \texttt{TRICERATOPS} supports the planetary interpretation of this outer signal.

\section{Prospects}
\label{sec:prospects_appendix}

\subsection{Prospects for mass determination}
\label{subsec:prospects_appendix_mass}

Accurate planetary masses are required to determine the bulk compositions of TOI-2079\,b and c and assess their potential for further characterization. We estimated the radial-velocity (RV) follow-up effort required using the publicly available RAVEX code\footnote{\url{https://github.com/franpoz/ravex}} \citep[RAdial VElocity eXplorer;][]{ravex2026}, following the general methodology described by \cite{barkaoui2025}, \cite{2026MNRAS.549ag183M}, and \cite{2026MNRAS.548ag713P}. We injected nominal masses derived from \texttt{SPRIGHT} \citep{Parviainen2024} of 2.4\,M$_{\oplus}$ and 5.0\,M$_{\oplus}$ for planets b and c, respectively. The orbital periods and phases were fixed to the photometric solution provided in Table~\ref{tab:parameters}.

We assumed observations with the MAROON-X spectrograph \citep{seifahrt2020}. For an exposure time of 1800\,s, the MAROON-X exposure-time calculator predicts signal-to-noise ratios of approximately 190 and 102 in the red and blue arms, corresponding to internal RV uncertainties of 0.91 and 1.11\,m~s$^{-1}$, respectively. We additionally adopted a conservative stellar-jitter term of 3\,m~s$^{-1}$ \citep{ruh2024}, treated as common to the simultaneous measurements from both arms but independent between observing epochs. For each number of observing epochs (N), we generated 100 synthetic campaigns by randomly sampling dates within the visibility windows and adding Gaussian noise according to the resulting covariance model. Each epoch therefore provides simultaneous red- and blue-arm RV measurements. The mass precision is defined as $\sigma_M/M_{\rm rec}$, where $M_{\rm rec}$ is the recovered planetary mass.

\begin{figure}
    \centering
    \includegraphics[width=\linewidth]{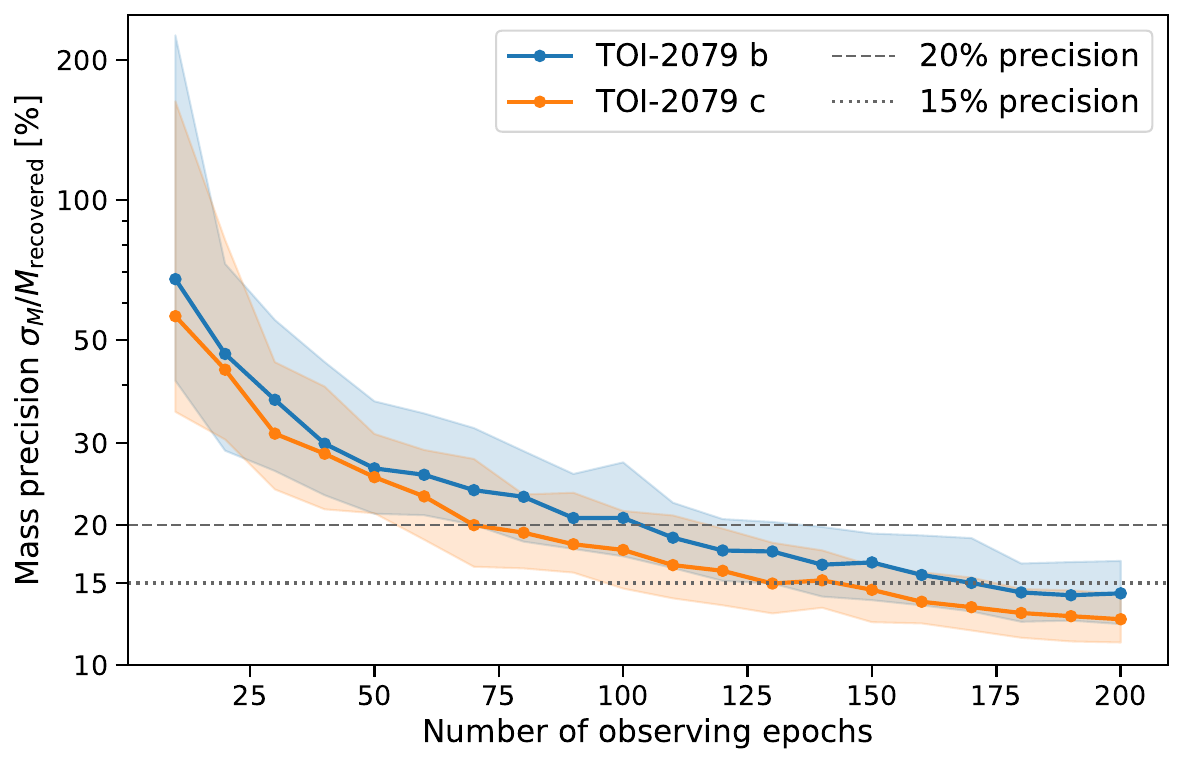}
    \caption{Expected precision of the recovered masses of TOI-2079\,b and c as a function of the number of MAROON-X observing epochs (N). Each epoch provides simultaneous RV measurements from the red and blue arms. Solid lines indicate median relative mass uncertainty, defined as $\sigma_M/M_{\rm rec}$, obtained from 100 simulated campaigns at each N. Shaded regions denote the 16th--84th percentile ranges, and the horizontal lines indicate reference precisions of 20 and 15\%.
}
    \label{fig:rvs}
\end{figure}

The resulting mass-precision curves are shown in Fig.~\ref{fig:rvs}. TOI-2079\,c reaches a median precision of approximately 20\% after 70 epochs, although the precision for TOI-2079\,b is still about 24\% at that point. A campaign of approximately 110 epochs, corresponding to about 55\,h of on-source time, yields expected precisions of approximately 19\% and 16\% for planets b and c, respectively. Increasing the campaign to approximately 170--180 epochs (85--90\,h) would improve these values to about 15 and 13\%, which, along with the current radius precision better than 4\% for both planets (see Table~\ref{tab:parameters}), would allow for reliable interior models \citep[see, e.g.,][]{plot2024,biceps2024}.

Nevertheless, the simulations do not explicitly model time-correlated stellar activity, and the estimated observing requirements should therefore be regarded as indicative planning values. In practice, time-correlated activity signals from the host star can introduce additional noise and systematic errors into radial-velocity measurements, potentially biasing the recovered planetary masses or degrading achievable precision, especially for the low-amplitude signals expected for TOI-2079\,b and c. 

\subsection{Prospects for atmospheric/surface characterization}
\label{subsec:prospects_appendix_atm}

The evolutionary stage of the TOI-2079 system, together with the location of TOI-2079\,c near the Radius Valley, makes both planets attractive targets for atmospheric characterization. Future observations may probe the presence of a secondary atmosphere, an escaping upper atmosphere, or reveal a bare-rock surface, thereby providing complementary constraints on the atmospheric evolution of rocky planets \citep{Gkouvelis2025} and on the transport processes regulating the helium abundance supplied to the escaping wind \citep{Gkouvelis2026He}.

\subsection{Atmosphere/bare rock surface}
\label{subsec:Lower atmosphere}

The Transmission Spectroscopy Metric (TSM) and Emission Spectroscopy Metric (ESM), as defined by \cite{Kempton2018}, provide hints about the feasibility of atmospheric/surface characterization and the most suitable observing strategies, particularly with the James Webb Space Telescope (JWST). With a TSM of $9 \pm 2$ and an ESM of $5.5 \pm 0.5$ (see Table \ref{tab:parameters}), TOI-2079 b ranks near or slightly below the thresholds delimiting the top quintile of the most favorable rocky planets for transmission and emission spectroscopy (thresholds of 10 and 7.5, respectively). TOI-2079 c is more challenging to characterize in emission due to its lower temperature. In transmission, however, its prospects are more favorable, with a TSM comparable to those of other similar-sized planets with detected atmospheric features, although well below the threshold of 90 for small sub-Neptunes.

To further assess the prospects for characterizing TOI-2079\,b and explore suitable observing strategies, we simulated JWST transmission and emission spectra under different atmospheric and surface scenarios.
For the atmospheric scenarios, we considered secondary gaseous envelopes composed of 100\% H$_2$O or 100\% CO$_2$, as well as the less likely possibility of a retained primordial atmosphere with up to 100$\times$ solar metallicity and haze. The synthetic spectra were generated with \texttt{TauREx} \citep{Waldmann2015a,Waldmann2015b,Al-Refaie2021} using the latest molecular cross sections available from the ExoMol database \citep{Chubb2021,Polyansky2018,Chubb2024,Yurchenko2026} and temperature-pressure profiles from the public grid by \cite{Samland2017}. For the 100$\times$ solar primordial atmosphere, thermochemical equilibrium was calculated with \texttt{GGchem} \citep{Woitke2018}, while haze opacity was modeled using a parameterized Mie scattering prescription \citep{Lee2013}.
For the bare-rock scenarios, we considered metal-rich, ultramafic and granitoid surface compositions as defined by \cite{Gkouvelis2025surface}, based on laboratory-measured reflectance and emissivity spectra from the USGS  and BUG spectral libraries \citep{Clark2007,Kokaly2017,Gkouvelis_et_al_2026}.

\begin{figure}
    \centering
    \includegraphics[width=\linewidth]{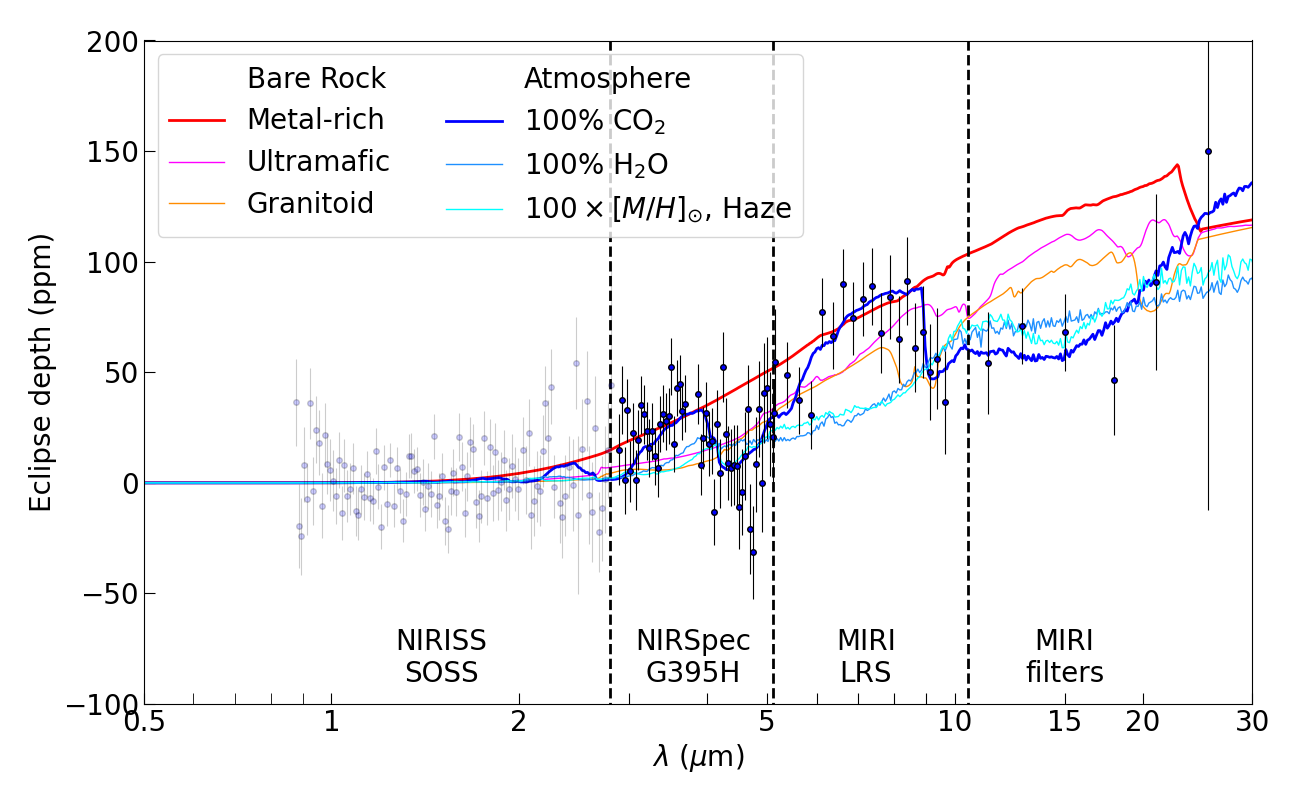}
    \includegraphics[width=\linewidth]{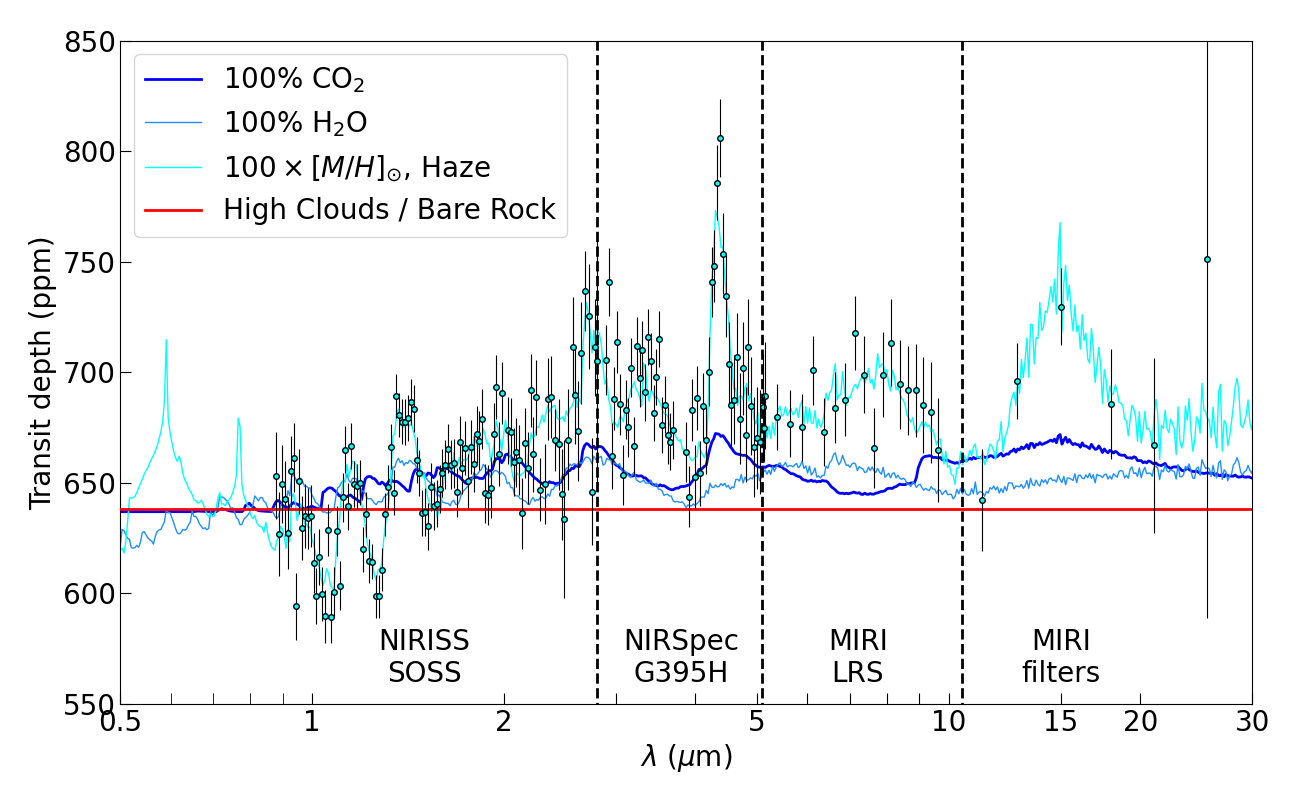}
    \caption{Modelled emission (top) and transmission (bottom) spectra of TOI-2079 b for three atmospheric compositions (100\% CO$_2$ in blue, 100\% H$_2$O in light blue, and 100$\times$ solar metallicity with haze in cyan) and three bare-surface compositions (metal-rich in red, ultramafic in magenta, and granitoid in orange). Synthetic observations (points with error bars) correspond to 16 visits using JWST NIRISS/SOSS, NIRSpec/G395H, MIRI/LRS, and the MIRI photometric filters F1130W, F1280W, F1500W, F1800W, F2100W and F2550W.}
    \label{fig:jwst_spectra}
\end{figure}

We then simulated the corresponding transmission and emission spectra as would be observed with JWST NIRISS/SOSS, NIRSpec/G395H, MIRI/LRS Slitless, and MIRI photometers (Figure \ref{fig:jwst_spectra}) using \texttt{ExoTETHyS} \citep{Morello2021}, following the same procedure as in our previous work (e.g., \citealp{Goffo2024,Lacedelli2025,2026MNRAS.548ag713P}). The error bars for the MIRI filters were independently derived using the JWST Exposure Time Calculator \citep{Pontoppidan2016}.
We note that $\gtrsim$10 visits are required to robustly detect atmospheric or surface features, even under optimistic scenarios. A secondary atmosphere produces very small features in transmission ($\sim$10 ppm), making it hardly distinguishable from a naked surface or a high-cloud atmosphere. An H$_2$--He-dominated atmosphere can produce order-of-magnitude larger features, but this scenario is unlikely for such a small planet orbiting an M-dwarf.
Emission features are more readily detectable at MIRI wavelengths, where simulated surface emission systematically exceeds atmospheric emission at $\sim$15\,$\mu$m under all simulated scenarios. This flux contrast motivates the single-filter observing strategy of the Rocky World DDT program \citep{Redfield2024}, although multi-color or full phase-curve observations are ultimately needed to break remaining degeneracies \citep{Gkouvelis2025RNAAS,Gillon2026}. Among the atmospheric scenarios, a Venus-like CO$_2$ composition produces distinct spectroscopic features detectable with MIRI/LRS and/or NIRSpec/G395H.

\subsection{Upper-atmosphere prospects}
\label{subsec:Upper atmosphere}

The estimated age and H$\alpha$ and Ly$\alpha$ activity of TOI-2079 motivate follow-up observations of atmospheric escape. The recent detection of escaping He\,\textsc{I} from the rocky planet LHS~1140\,b \citep{Cherubim2026} demonstrates the potential of the 10830\,\AA\ triplet to probe escape from small planets. We therefore computed synthetic He\,\textsc{I} 10830\,\AA\ transmission spectra for both TOI-2079 planets over a grid of thermospheric maximum temperatures ($T_{\rm max}$) and atmospheric mass-loss rates ($\dot{M}$), and estimated the number of CARMENES transits required for a $3\sigma$ detection. The models span conditions considered plausible for irradiated rocky planets \citep{Gkouvelis2026}.

\begin{figure}
\centering
\includegraphics[scale=0.35, trim=1.5cm 0cm 0cm 0cm, clip]{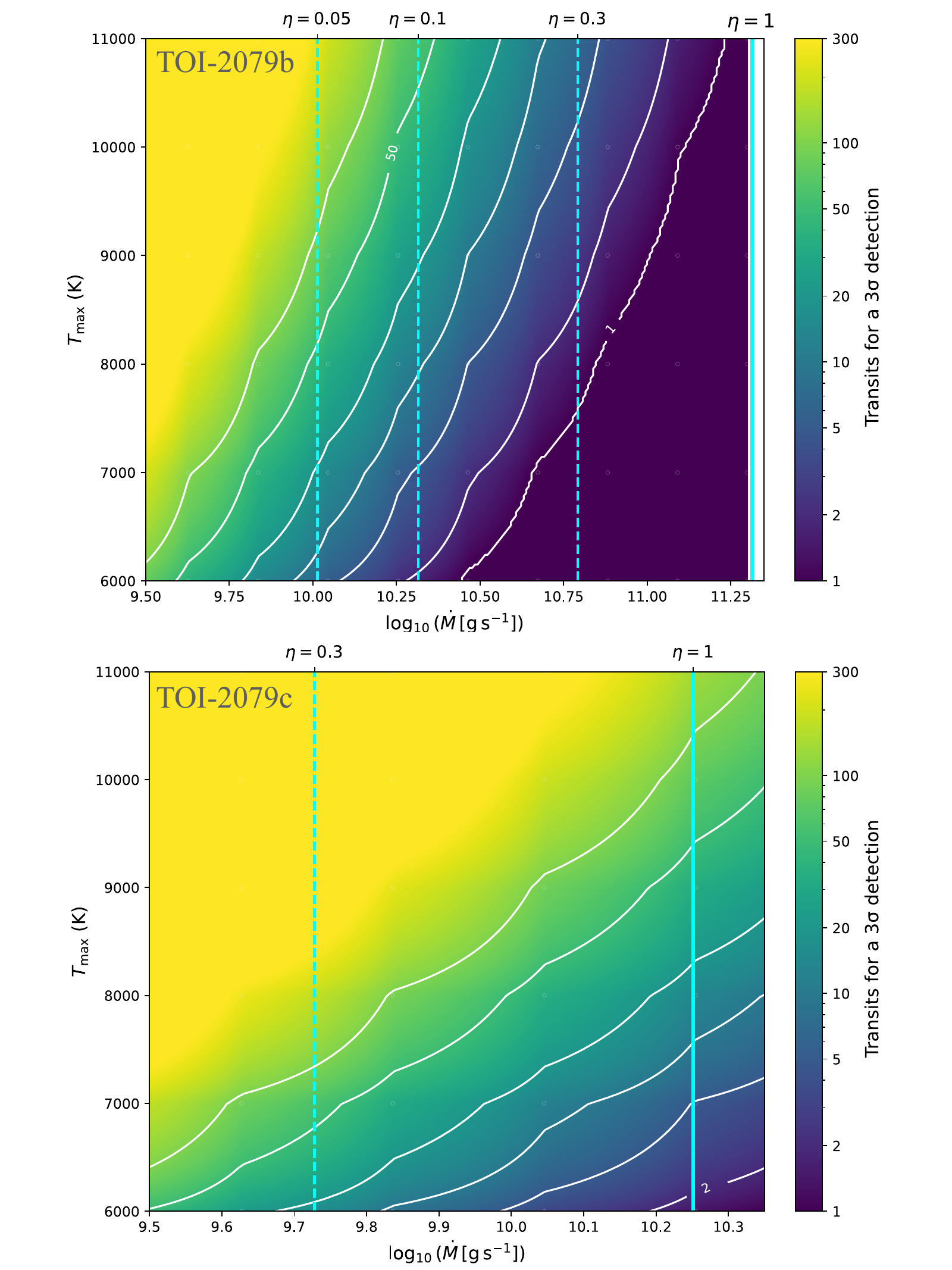}
\caption{Predicted detectability of the He\,\textsc{i} 10830\,\AA\ triplet for planet b (top) and c (bottom) as a function of the thermospheric maximum temperature, $T_{\rm max}$, and atmospheric mass-loss rate, $\dot{M}$. The color scale indicates the estimated number of CARMENES transit observations required to achieve a $3\sigma$ detection. Cyan vertical lines indicate the energy-limited mass-loss rates corresponding to heating efficiencies, $\eta$. Values of $\eta$ between $\sim0.01$ and $\sim0.6$ are commonly inferred for hydrodynamic atmospheric escape, while $\eta=1$ represents the theoretical upper limit.}
\label{fig:Helium_escape}
\end{figure}

Figure~\ref{fig:Helium_escape} shows the predicted number of transits required for detection. The cyan lines mark energy-limited mass-loss rates for representative heating efficiencies. For TOI-2079\,b, rates corresponding to $\eta \sim 0.01$--$0.6$ overlap with model predictions requiring only a few transits. Detectable absorption is also possible for TOI-2079\,c over part of the modeled parameter space, despite its lower irradiation. These estimates depend on the assumed escape conditions and do not establish that either planet currently hosts an escaping atmosphere. They indicate that He\,\textsc{I} observations could test for atmospheric escape in this system, including planets at the Radius Valley.

\bibliography{sample701}{}
\bibliographystyle{aasjournalv7}

\end{document}